\documentclass[english,american,superscriptaddress,prd,a4paper,nofootinbib,11pt,eqsecnum,secnumarabic]{revtex4-2}
\usepackage{multirow,slashed,float,graphicx,caption,subcaption,hyperref,color}
\usepackage[dvipsnames]{xcolor}
\usepackage[utf8]{inputenc}
\usepackage{braket}
\hypersetup{colorlinks=true, linkcolor=BurntOrange, citecolor=BurntOrange,filecolor=BurntOrange, urlcolor=BurntOrange}
\usepackage{mathtools}
\usepackage{amsmath}
\usepackage{ulem}
\newcommand{\be}{\begin{equation}}
\newcommand{\ee}{\end{equation}}
\newcommand{\bea}{\begin{eqnarray}}
\newcommand{\eea}{\end{eqnarray}}

\def\bsp#1\esp{\begin{split}#1\end{split}}

\usepackage{epsfig}
\usepackage{amsbsy}
\usepackage{varioref}
\usepackage{pifont}
\usepackage{amsmath,amssymb}
\usepackage{wrapfig}

\usepackage{mdframed}
\usepackage[export]{adjustbox}

\begin{document}

%%%%%%%%%%%%%%%%%%%%%%%%%%%%%%%%%%%%%%%%%%%%%%%%%%%%%%%%%%%%%%%%%%%%%%%%%%%
\title{Scattering amplitudes in the Randall-Sundrum model\\ with brane-localized curvature terms}

\author{R. Sekhar Chivukula}
\email{rschivukula@ucsd.edu}
\affiliation{Department of Physics, University of California, San Diego, 9500 Gilman Drive, La Jolla, CA-92093, USA
}
\author{Kirtimaan A. Mohan}
\email{kamohan@msu.edu}
\affiliation{Department of Physics and Astronomy, 
Michigan State University\\
567 Wilson Road, East Lansing, MI-48824, USA}
\author{Dipan~Sengupta}
\email{dipan.sengupta@unsw.edu.au}
\affiliation{Sydney Consortium for Particle Physics and Cosmology, School of Physics,\\ The University of New South Wales, Sydney NSW 2052, Australia}
\affiliation{ARC Centre of Excellence for Dark Matter Particle Physics\\
Australia}
\author{Elizabeth H. Simmons}
\email{ehsimmons@ucsd.edu}
\affiliation{Department of Physics, University of California, San Diego, 9500 Gilman Drive, La Jolla, CA-92093, USA
}
\author{Xing Wang}
\email{xiw006@physics.ucsd.edu}
\affiliation{Department of Physics, University of California, San Diego, 9500 Gilman Drive, La Jolla, CA-92093, USA
}

%%%%%%%%%%%%%%%%%%%%%%%%%%%%%%%%%%%%%%%%%%%%%%%%%%%%%%%%%%%%%%%%%%%%%%%%%%%
\begin{abstract}
In this paper we investigate the scattering amplitudes of spin-2 Kaluza-Klein (KK) states in Randall-Sundrum models with brane-localized curvature terms. We show that the presence of brane-localized curvature interactions modifies the properties of (4D) scalar fluctuations of the metric, resulting in scattering amplitudes of the massive spin-2 KK states which grow as ${\cal O}(s^3)$ instead of ${\cal O}(s)$. We discuss the constraints on the size of the brane-localized curvature interactions based on the consistency of the Sturm-Liouville mode systems of the  spin-2 and spin-0 metric fluctuations. We connect the properties of the scattering amplitudes to the diffeomorphism invariance of the compactified KK theory with brane-localized curvature interactions. We verify that the scattering amplitudes involving brane-localized external sources (matter) are diffeomorphism-invariant, but show that those for matter localized at an arbitrary point in the bulk are not. We demonstrate that, in Feynman gauge, the spin-0 Goldstone bosons corresponding to helicity-0 states of the massive spin-2 KK bosons behave as a tower of Galileons, and that it is their interactions that produce the high-energy behavior of the scattering amplitudes. We also outline the correspondence between our results and those in the Dvali-Gabadadze-Porrati (DGP) model. In an appendix we discuss the analogous issue in extra-dimensional gauge theory, and show that the presence of a brane-localized gauge kinetic-energy term does not change the high-energy behavior of corresponding KK vector boson scattering amplitudes.

\end{abstract}

\maketitle
{
  \hypersetup{linkcolor=blue}
  \tableofcontents
}
\newpage
\flushbottom

%%%%%%%%%%%%%%%%%%%%%%%%%%%%%%%%%%%%%%%%%%%%%%%%%%%%%%
%%%%%%%%%%%%%%%%%%%%%%%%%%%%%%%%%%%%%%%%%%%%%%%%%%%%%%
\section{Introduction}

Theories incorporating compact extra-dimensions offer a possible solution to the Standard Model hierarchy problem, as well as a viable pathway to solve a large number of other puzzles including the dark matter and flavor problems.
%\cite{} {\bf REFERENCES NEEDED}. 
Since there are no phenomenologically viable UV-complete extra-dimensional theories, these models must be understood as effective field theories. It is therefore crucial that we assess the validity of these theories in terms of the maximum energy scale (the ``UV cutoff" of the effective theory) to which they provide a useful description of physics.

The gravitational sector of extra-dimensional theories
%with an extra compact spatial dimension
includes (in addition to a massless graviton) an interacting tower of massive spin-2 Kaluza-Klein states \cite{Kaluza:1921tu,Klein:1926tv}.   
The tree-level scattering amplitudes of these states grow potentially as fast as ${\cal O}\left(s^5/{\Lambda^2 m^8_{KK}}\right)$ \cite{Arkani-Hamed:2002bjr,Arkani-Hamed:2003roe}, where $s$ is the center-of-mass scattering energy-squared, $m_{KK}$ is the mass of the Kaluza-Klein spin-2 states involved, and $\Lambda$ is an intrinsic scale associated with the compactified gravitational theory (specifically, the Planck mass in a theory in which the extra-dimension is flat, or that scale suitably ``warped down" in the case of a Randall-Sundrum model \cite{Randall:1999ee,Randall:1999vf}). Scattering amplitudes growing this quickly would result in a violation of unitarity and an upper bound on the scale of validity of the theory at an energy {\it inversely proportional} to the mass of the Kaluza-Klein states, and much lower than the intrinsic gravitational scale $\Lambda$.\footnote{This is  the expected growth in theories of massive Fierz-Pauli gravity, where general relativity is modified by adding an explicit mass term \cite{deRham:2014zqa,Hinterbichler:2011tt}.}

Recently, through detailed calculations \cite{SekharChivukula:2019yul,SekharChivukula:2019qih,Chivukula:2020hvi,Chivukula:2021xod} (see also \cite{Bonifacio:2019ioc}) of the massive spin-2 scattering amplitudes in compactified five-dimensional gravity, it has been shown that although the contributions to the scattering amplitudes from individual Feynman diagrams do indeed grow as fast as ${\cal O}(s^5/\Lambda^2 m^8_{KK})$, there are typically cancellations between different contributions to the scattering amplitudes which reduce the overall rate of growth to ${\cal O}(s/\Lambda^2)$ -- the result which would have been naively expected in consistent four-dimensional gravitational theories. It has further been shown that the cancellations observed are the result of ``sum rules" \cite{SekharChivukula:2019qih,Bonifacio:2019ioc} relating the masses and couplings of the spin-2 modes, which hold in both flat and warped extra dimensions, arise from the underlying diffeomorphism symmetries of the underlying theory \cite{Hang:2021fmp,Hang:2022rjp,Chivukula:2023qrt,Hang:2024uny}, and persist in models in which the radion is massive and the size of the extra dimension is stabilized via a Goldberger-Wise \cite{Goldberger:1999uk,Goldberger:1999un} mechanism \cite{Chivukula:2021xod,Chivukula:2022tla,Chivukula:2022kju}. These results have been generalized to consider the interactions of the compactified gravitational theory with matter, either in the bulk or localized on the ``branes" bounding the extra dimension \cite{deGiorgi:2020qlg,Chivukula:2023sua}.

In this paper we extend the work described above to consider the case in which the Randall-Sundrum gravitational theory includes four-dimensional brane-localized curvature terms in addition to the bulk gravity Einstein-Hilbert term.\footnote{Per usual in Randall-Sundrum models, appropriate bulk and brane cosmological constants must also be included in order to arrive at the desired warped background geometry.} While the addition of the brane-localized curvature terms does not change the background geometry (since the branes are Ricci-flat four-dimensional surfaces), the brane-localized curvature terms modify the properties of the fluctuations of the metric \cite{George:2011sw,Dillon:2016bsb}.\footnote{By contrast, these complications do not arise in a extra-dimensional gauge-theory with brane-localized gauge-kinetic terms, as we show in Appendix \ref{sec:appendix}.}
The corresponding changes to the the wavefunctions  of the massive spin-2 Kaluza-Klein modes (specifically, through modifications in the boundary conditions) modify the masses of these modes \cite{Davoudiasl:2003zt,Shtanov:2003um,Carena:2005gq,Bao:2005bv,George:2011sw,Dillon:2016bsb} and therefore of the couplings between them.

Our analysis of models with brane-curvature interactions extends those presented previously in the literature.
We discuss constraints \cite{Davoudiasl:2003zt,Luty:2003vm,George:2011sw,Miao:2023mui} on the size of the brane-localized curvature interactions based on the consistency of the Sturm-Liouville mode systems of the  spin-2 and spin-0 metric fluctuations.  We further show that the spin-2 KK scattering amplitudes in the presence of  brane-localized curvature interactions no longer display all of the cancellations described above; the overall scattering amplitudes instead grow like ${\cal O}(s^3/\Lambda^6_3)$, where $\Lambda_3$ is a scale which depends on the size of the brane-localized curvature terms present. 
%This amplitude behavior is unlike that of gauge fields, or fermions or scalars localized on the brane interacting with gravity, where the  scattering amplitudes grow no faster than $s/\Lambda^{2}$. {\color{Red} MARK!}

The behavior of the scattering amplitudes of massive spin-2 states is opaque in unitary gauge, in which the polarization vectors of the helicity-0 components, the internal spin-2 massive propagators, and the interaction vertices all grow like energy-squared at high energies. Instead, utilizing the diffeomorphism symmetries of the background geometry \cite{Dolan:1983aa,Cho:1992rq,Chivukula:2023qrt}, we analyze the theory in Feynman gauge \cite{Hang:2021fmp,Hang:2022rjp,Chivukula:2023qrt,Hang:2024uny} which includes (in addition to the massive spin-2 states) unphysical Goldstone vector and scalar states which are degenerate \cite{Lim:2005rc,Lim:2007fy,Lim:2008hi,Chivukula:2022kju} with the spin-2 KK modes. In particular, the Feynman gauge scattering amplitudes for helicity-0 and helicity-1 KK gravitons are related by an Goldstone boson equivalence theorem \cite{Hang:2021fmp,Hang:2022rjp,Chivukula:2023qrt,Hang:2024uny} to the corresponding amplitudes involving the unphysical vector and scalar Goldstone modes, which can then be understood easily by power-counting.

In Randall-Sundrum theories without brane-localized curvature terms, the diffeomorphism symmetries of the background (the subset of the $AdS_5$ diffeomorphism symmetries which respect the presence of the TeV and Planck branes bounding the internal space) are simply related to the transformation properties of the metric fluctuations \cite{Hang:2021fmp,Hang:2022rjp,Chivukula:2023qrt,Hang:2024uny}, and the only residual growth in the amplitudes arises from the energy-squared growth in the gravitational interactions themselves. 

In this work we show that the situation changes once brane-localized curvature terms are present. While the diffeomorphism symmetries of the background geometry are unchanged by the presence of brane-localized curvature, their action on the properly defined metric fluctuation fields is different.  Having properly identified the action of diffeomorphisms on the metric fluctuations, we can define and analyze the theory in Feynman gauge.  The change in the properties of the scalar metric fluctuations in the presence of brane-localized curvature terms requires a change in how these fluctuations are parameterized \cite{George:2011sw,Dillon:2016bsb} and leads to new Galileon interactions \cite{Dvali:2000hr,Luty:2003vm,Nicolis:2008in,deRham:2010ik,Hinterbichler:2011tt,deRham:2014zqa} for the tower of Goldstone bosons corresponding to the helicity-0 modes of the massive spin-2 fields. We show that  it is these Galileon interactions that produce the $O(s^3/\Lambda^6_3)$ high-energy behavior of the spin-2 KK scattering amplitudes.

Additionally, following \cite{Lim:2008hi},
we explicitly verify that the scattering amplitudes involving brane-localized external sources (matter) are independent of the gauge chosen in the gravitational sector and show that this gauge-invariance relies crucially on the proper identification of the diffeomorphism invariance of the theory. We also show that the scattering amplitudes for matter localized at an arbitrary point in the bulk are not gauge-invariant. This behavior is consistent with the anomalous growth of scattering amplitudes of bulk-localized matter found  in \cite{Chivukula:2023sua}.

Finally, we also outline the correspondence between our results and those in the Dvali-Gabadadze-Porrati (DGP) model \cite{Dvali:2000hr,Luty:2003vm}, which also includes both bulk and brane-localized curvature interactions, but with a semi-infinite internal space. In the DGP model the semi-infinite extra dimension leads to a resonance graviton continuum of states.  We show how our analysis of the model with a finite extra-dimension and a discrete gravity KK spectrum can be interpreted in the limit in which one brane is moved to infinity.  We recover the results of \cite{Dvali:2000hr,Luty:2003vm}, and show that  the interpretation of the high-energy scattering amplitudes in terms of the Goldstone boson equivalence theorem reproduces the high-energy behavior of the scattering amplitudes in the DGP model described in \cite{Dvali:2000hr,Luty:2003vm}.

The outline of the paper is as follows. In section \ref{sec:RSmodel} we establish notation, introduce the parametrization used to describe metric fluctuations, and show how the parametrization introduced allows one to derive a convenient form for the quadratic action for the metric fluctuations. In section \ref{sec:modeexpansion} we introduce the Kaluza-Klein mode expansion appropriate to this situation and describe the hidden supersymmetric structure of the corresponding Sturm-Liouville systems. In section \ref{sec:5Ddiffeomorphism} we investigate the diffeomorphism symmetry of the system, and demonstrate that the action of the diffeomorphisms on the metric fluctuations differs from the case without brane-localized curvature interactions. In section \ref{sec:GF} we describe $R_\xi$ gauge-fixing of the gravitational sector and demonstrate the gauge-independence of the scattering amplitudes involving brane-localized external sources. Section \ref{sec:Scattering} presents our primary results for the behavior of the scattering amplitudes of the massive spin-2 KK states, demonstrates the ${\cal O}(s^3)$ growth of these amplitudes, investigates the relationship between the strong coupling scales $\Lambda_3$ and $\Lambda$, and the masses of the spin-2 states. We also show in this section that the ${\cal O}(s^3)$ growth of the scattering amplitudes can be understood using the Goldstone boson equivalence theorem in terms of Galileon interactions of the tower of spin-0 states present in `t-Hooft-Feynman gauge. Section \ref{sec:DGP} investigates the correspondence between our results and the DGP model. The final section~\ref{sec:conclusion} presents our conclusions. In the appendix~\ref{sec:appendix} we 
consider compactified 5-dimensional gauge theory in the presence of brane gauge kinetic energy terms, and show that no change in the high energy behavior of the scattering amplitudes of the massive spin-1 modes results from the inclusion of brane-localized kinetic terms.

%%%%%%%%%%%%%%%%%%%%%%%%%%%%%%%%%%%%%%%%%%%%%%%%%%%%%%

\section{Randall-Sundrum Model with brane-localized curvature terms}

\label{sec:RSmodel}

The line element of the Randall-Sundrum model (RS1) is written, in conformal coordinates  $(x^{\mu},z)$, as
\begin{equation}
    ds^2 = e^{2A(z)}(\eta_{\mu\nu}dx^\mu dx^\nu - dz^2),
\end{equation}
where the background 4D Minkowski metric $\eta_{\mu\nu}\equiv{\rm diag}(+1,-1,-1,-1)$ is used to raise and lower Greek indices, and $z$ lies in the interval $z_1=1/k \le z \le z_2=e^{k\pi r_c}/k$, where $k$ is the AdS curvature and $r_c$ is the RS1 ``compactification radius," a measure of the size of the internal dimension. The warp factor $A(z)$ is given by, 
\begin{equation}
   A(z) = -\ln(kz)~.
\end{equation}
The 5D gravitational Lagrangian, augmented by the brane-localized curvature terms on the UV brane $(z=z_1)$ and IR brane $(z=z_2)$, is given by \cite{Shtanov:2003um,Carena:2005gq,Bao:2005bv,George:2011sw,Dillon:2016bsb} 
\begin{equation}
\aligned
    \mathcal{L} ~=~& M_5^3 \left[ \sqrt{G}\,R^{(5)} + \delta(z-z_1^+)\,\gamma_1\, e^A\sqrt{-g}\,R^{(4)} + \delta(z-z_2^-)\,\gamma_2\, e^A\sqrt{-g}\,R^{(4)}\right] \\
    &~+ \mathcal{L}_{\rm CC}+\Delta \mathcal{L} + \mathcal{L}_{\rm GF},
\endaligned
\end{equation}
where $R^{(5)}$ is the 5D Ricci scalar constructed from the 5D metric $G_{MN}$, and $R^{(4)}$ is the brane-localized 4D Ricci scalar constructed from the induced 4D metric $g_{\mu\nu}$ on the branes. Here we use the notation $z^\pm \equiv \lim_{\varepsilon \to 0^+} z\pm \varepsilon$, $M_5$ is the 5D Planck mass, and $\gamma_{1,2}$ are the parameters that characterize the strength of the brane-localized curvature terms and have the mass dimension of $[{\rm mass}]^{-1}$. The $\mathcal{L}_{\rm CC}$ include the cosmological constant terms, and  $\Delta\mathcal{L}$ is a total derivative term required for a well-defined variational principle for the action. $\mathcal{L}_{\rm GF}$ is gauge fixing term that will be specified later in Sec.~\ref{sec:GF}. In addition we have defined the Dirac delta function in the interval picture such that
\begin{equation}
    \int_{z_1}^{z>z_1}dz~\delta(z-z_1^+) = \int_{z<z_2}^{z_2}dz~\delta(z-z_2^-) = 1.
\end{equation}

We parametrize the metric, including fluctuations, as
\begin{equation}
    G_{MN} = e^{2A(z)}\left(\eta_{MN} + \kappa H_{MN}\right),
\end{equation}
with five-dimensional coordinate labels $M,N=0, 1, 2, 3, 4$. The metric fluctuations in turn are given by \cite{George:2011sw,Dillon:2016bsb},\footnote{In refs. \cite{George:2011sw,Dillon:2016bsb} there are three scalar metric fluctuation fields introduced, which are labeled $P_{1,2,3}(x^\alpha,z)$. The equations of motion (Einstein field equations) enforce $P_1\equiv P_2$, and the corresponding single physical field we label here $\varphi$.  $P_3$, which is an unphysical auxiliary field introduced to facilitate the diagonalizing the mode fluctuations, is here labeled $\Delta$. This form of metric is inspired by cosmological considerations of brane-world models, see for example \cite{Deffayet:2002fn,Bridgman:2001mc}.}
\begin{equation}
    H_{MN} = \begin{pmatrix}
        h_{\mu\nu} - \dfrac{1}{\sqrt{6}}\eta_{\mu\nu}\left(\varphi -A' \Delta'\right) +\dfrac{1}{\sqrt{6}}\partial_\mu\partial_\nu \Delta & \dfrac{\epsilon(z)}{\sqrt{2}}A_\mu \\
        \dfrac{\epsilon(z)}{\sqrt{2}}A_\nu & -\sqrt{\dfrac{2}{3}}\left(\varphi + \dfrac{1}{2}\Delta'' + \dfrac{1}{2}A'\Delta'\right)
    \end{pmatrix}~,
    \label{eq:background-metric}
\end{equation}
where the Lorentz indices are $\mu=0,1,2,3$, $h_{\mu\nu}(x^\mu,z)$ is the spin-2 tensor field, $A_\mu(x^\mu,z)$ is the spin-1 vector field, $\varphi(x^\mu,z)$ and $\Delta(x^\mu,z)$ are spin-0 scalar fields, and primes denote derivatives with respect to the internal coordinate $z$. Here $\epsilon(z)$ is a rectangular function,
\begin{equation}
    \epsilon(z)=\begin{cases}
        1,\quad z_1<z<z_2 \\
        0,\quad z=z_{1,2}
    \end{cases},\qquad \epsilon'(z) = \delta(z-z_1^+)-\delta(z-z_2^-),
\end{equation}
which imposes the ``straight gauge" conditions \cite{Carena:2005gq} that 
\begin{align}
 G_{\mu 4}(z_{1,2}) \equiv 0~
 \label{eq:straight-gauge}
\end{align}
at the boundaries of the internal space.

The parametrization of the scalar fluctuations using the fields $\varphi(x^\mu,z)$ and $\Delta(x^\mu,z)$ given in Eq.~(\ref{eq:background-metric}), which was introduced in \cite{Deffayet:2002fn,Bridgman:2001mc,George:2011sw,Dillon:2016bsb}, may be unfamiliar. At first sight it would seem that the introduction of the field $\Delta(x^\mu,z)$ is redundant for two separate reasons. First, $\Delta$ appears in the bulk metric as a pure gauge degree of freedom. Under an infinitesimal coordinate transformation,
\begin{equation}
    x^M\mapsto \overline{x}^M=x^M+\xi^M~,
    \label{eq:linear-diffeomorphisms}
\end{equation}
one can show (see Eqs.~(\ref{eq:diffeomorphism-begin}) - (\ref{eq:diffeomorphism-end})) that the choice 
\begin{equation}
    \xi_\mu = \dfrac{1}{2\sqrt{6}} \partial_\mu\Delta, \ \ \text{and} \ \ \xi^5=\dfrac{1}{2\sqrt{6}} \Delta'~,
\end{equation}
would apparently remove the field from the metric entirely, setting $\Delta\equiv 0$. Furthermore, the fields $\varphi$ and $\Delta$ should ultimately describe a single massless physical degree of freedom, the radion, and thus cannot be mutually independent.\footnote{In the context of the DGP model \cite{Dvali:2000hr}, the field $\Delta$ is gauge-equivalent to the strongly-coupled Galileon scalar mode \cite{Luty:2003vm} $\Pi$ found in the fluctuations  of ADM \cite{Arnowitt:1962hi}  ``lapse" and ``shift", as we show directly in Sec.~\ref{sec:DGP}.}

 However, following \cite{George:2011sw,Dillon:2016bsb}, we constrain $\Delta(x^\mu, z)$ by eliminating the brane-localized kinetic-energy mixing terms that arise between the tensor and scalar fields. In particular, expanding the Lagrangian to quadratic order we find the brane mixing terms
\begin{equation}
    \mathcal{L}_{h\mbox{-}\varphi/\Delta} = \sum_{i=1,2} \delta_i\,e^{3A}\left(\partial_\mu\partial_\nu h^{\mu\nu}-\Box h\right)\left[2\gamma_i\,\varphi + \left(\theta_i-2A'\gamma_i\right)\Delta'\right] + {\rm non\mbox{-}}\partial_\mu{\rm\, terms},
    \label{eq:mode-mixing}
\end{equation}
where $\delta_1 \equiv \delta(z-z_1^+)$, $\delta_2 \equiv \delta(z-z_2^-)$, and $\theta_1=-\theta_2=1$. Note that in the absence of brane-localized curvature, $\gamma_i \equiv 0$, we can consistently take $\Delta\equiv 0$. In the presence of brane-localized curvature, however, the kinetic tensor-scalar mixing is instead eliminated by choosing,
\begin{equation}
    \Delta'(z=z_i) = -\dfrac{2\gamma_i}{\theta_i-2A'(z_i)\gamma_i}\varphi(z=z_i).
    \label{eq:P3}
\end{equation}
This condition fixes $\Delta'$ on the boundaries $z=z_i$ in terms of the value of the field $\varphi$ on the boundary. $\Delta(x^\mu,z)$ is therefore understood to be an auxiliary field which cannot be eliminated in analyzing the theory with brane-localized curvature terms.  The condition in Eq.~(\ref{eq:P3}) does not constrain the bulk value of $\Delta$, however, and we will introduce a convenient choice for the field $\Delta(x^\mu,z)$ in Sec.~\ref{subsec:Equivalance}.

Once we fix $\Delta'$ on the boundaries, the quadratic action can be written as,
\begin{equation}
    S_2 = \int d^4x~dz~e^{3A(z)}\left( \mathcal{L}_{h\mbox{-}h} + \mathcal{L}_{h\mbox{-}A} + \mathcal{L}_{h\mbox{-}\varphi} + \mathcal{L}_{A\mbox{-}A} + \mathcal{L}_{A\mbox{-}\varphi} + \mathcal{L}_{\varphi\mbox{-}\varphi} \right),
    \label{eq:RS1Lagrangian}
\end{equation}
with
\begin{align}
    \mathcal{L}_{h\mbox{-}h} =&~ h_{\mu\nu}\left[\vphantom{\frac{1}{4}}\right.\frac{1}{4}\left(\eta^{\mu\rho}\partial^\nu\partial^\sigma + \eta^{\mu\sigma}\partial^\nu\partial^\rho + \eta^{\nu\rho}\partial^\mu\partial^\sigma + \eta^{\nu\sigma}\partial^\mu\partial^\rho\right) \nonumber\\
    & \hspace{1cm}-\frac{1}{2}\left(\eta^{\mu\nu}\partial^\rho\partial^\sigma + \eta^{\rho\sigma}\partial^\mu\partial^\nu\right) \nonumber\\
    & \hspace{1cm}\left.- \frac{1}{4}\left(\eta^{\mu\rho}\eta^{\nu\sigma} + \eta^{\mu\sigma}\eta^{\nu\rho} - 2\eta^{\mu\nu}\eta^{\rho\sigma}\right)\Box \right] \,h_{\rho\sigma}\left(1+\sum_i\gamma_i\delta_i\right)\nonumber\\
    & \hspace{.5cm}- \frac{1}{4}\left(\eta^{\mu\rho}\eta^{\nu\sigma} + \eta^{\mu\sigma}\eta^{\nu\rho} - 2\eta^{\mu\nu}\eta^{\rho\sigma}\right)\left( Dh_{\mu\nu}\right)\left(Dh_{\rho\sigma}\right),\label{eq:RS1hh}\\
    \mathcal{L}_{h\mbox{-}A} =&~ -\dfrac{1}{\sqrt{2}}\partial_\mu hD^\dagger A_\mu + \left(Dh_{\mu\nu}\right)\left[\vphantom{\frac{1}{4}}\right.\left.\frac{1}{\sqrt{2}}\left(\eta^{\mu\rho}\partial^\nu + \eta^{\nu\rho}\partial^\mu - \eta^{\mu\nu}\partial^\rho\right)\right]A_\rho \nonumber\\
    &~+\sum_i\sqrt{2}\,\delta_i\theta_i h\,\partial_\mu A^\mu,\label{eq:RS1hA}\\
    \mathcal{L}_{A\mbox{-}A} =&~ A_{\mu}\left[\vphantom{\frac{1}{4}}\right.\left.-\frac{1}{2}\left(\partial^\mu\partial^\nu-\eta^{\mu\nu}\Box\right)\right]A_\nu,\label{eq:RS1AA}\\
    \mathcal{L}_{h\mbox{-}\varphi} =&~ \left(Dh\right)\left[\vphantom{\frac{1}{4}}\sqrt{\frac{3}{2}}\overline{D}^\dagger \right]\varphi,\label{eq:RS1hphi}\\
    \mathcal{L}_{A\mbox{-}\varphi} =&~ \varphi\left[\vphantom{\frac{1}{4}}\sqrt{3}\ \partial^\mu\overline{D} \vphantom{\frac{1}{4}}\right]A_{\mu} + \sum_i2\sqrt{3}\,\delta_i\theta_i\,\varphi\,\partial_\mu A^\mu,\label{eq:RS1Aphi}\\
    \mathcal{L}_{\varphi\mbox{-}\varphi} =&~ -\frac{1}{2}\varphi\Box\varphi \left(1+\sum_i\dfrac{\gamma_i\delta_i}{1-2A'\gamma_i\theta_i}\right) + \left(\overline{D}^\dagger\varphi\right)^2,\label{eq:RS1phiphi}
\end{align}
where the differential operators are defined as,
\begin{equation}
    D = \partial_z,\quad D^\dagger = -(\partial_z+3A'),\quad \overline{D} = \partial_z+A',\quad \overline{D}^\dagger=-(\partial_z+2A').
    \label{eq:D-definitions}
\end{equation}

Importantly, the quadratic Lagrangian is completely free of the field $\Delta$. Therefore, $\Delta$ is {\it not} an independent dynamical field. In fact, as would be expected since $\Delta$ appears nominally as a gauge degree of freedom, its value in the bulk is entirely unconstrained! Only its derivative at the boundaries is constrained by Eq.~(\ref{eq:P3}). Once we choose a convenient value of $\Delta$ consistent with the boundary constraint (in essence choosing a ``reference frame" in which to properly define our metric fluctuations), this field is determined in terms of the field $\varphi$. The interactions of $\Delta$ arising from its presence in the metric in Eq.~(\ref{eq:background-metric}), therefore, will give rise to new interactions of the scalar ($\varphi$) degrees of freedom in the gravitational sector. We will discuss these topics further from the symmetry point of view in Sec.~\ref{sec:GF} and consider the consequences of the additional interactions arising from the presence of $\Delta$ for spin-2 scattering amplitudes in Sec.~\ref{subsec:Equivalance}.

\section{Mode Expansion and Supersymmetric Structure}

\label{sec:modeexpansion}

Next we consider decomposing the five-dimensional metric-fluctuation fields $h_{\mu\nu}(x^\alpha,z)$, $A_\mu(x^\alpha,z)$, and $\varphi(x^\alpha,z)$, into four-dimensional Kaluza-Klein modes and determine the boundary conditions which must be satisfied by the corresponding mode functions. These decompositions, and the Sturm-Liouville systems associated with the corresponding mode wavefunctions, are most easily determined by considering the symmetry properties of the theory. In particular, the underlying five-dimensional diffeomorphism invariance of the non-compact  gravitational theory is broken by compactification to an infinite-dimensional Kac-Moody-like symmetry of the four-dimensional Kaluza-Klein theory, and the infinite tower of these symmetries is spontaneously broken to the residual four-dimensional Poincar{\`e} invariance \cite{Dolan:1983aa,Cho:1992rq}. The Goldstone bosons of the broken Kac-Moody symmetries are comprised of a tower of vector states associated with the spontaneously broken Kac-Moody symmetries related to four-dimensional translations, and a tower of scalar states associated with the spontaneous breaking of translations along the compact direction \cite{Dolan:1983aa,Cho:1992rq}. These Goldstone modes, with the exception of the lowest mass spin-0 radion state which remains in the spectrum, are ``eaten" by the massive spin-2 tower and correspond to the helicity-1 and helicity-0 of those Kaluza-Klein states.

The relationship between the Kaluza-Klein mode expansion and the underlying five-dimensional diffeomorphism invariance (as expressed in the infinite Kac-Moody symmetries of the four-dimensional Lagrangian) was first uncovered in the case of flat extra dimensions \cite{Dolan:1983aa,Cho:1992rq}, and subsequently extended to the case of Randall-Sundrum warped compactifications \cite{Lim:2005rc,Lim:2007fy,Lim:2008hi,Chivukula:2022kju,Chivukula:2023qrt}. In particular, the authors of \cite{Lim:2005rc,Lim:2007fy,Lim:2008hi} uncovered two ``hidden" $N=2$ quantum-mechanical supersymmetries (SUSY)\footnote{For a review, see \cite{Cooper:1994eh}.} relating the Sturm-Liouville systems of the spin-2, spin-1, and spin-0 wavefunctions. These hidden supersymmetries ensure that the towers of spin-1 and spin-0 modes are degenerate with the spin-2 modes,\footnote{This condition is (almost, aside from the identification of which towers have zero-modes) trivial in the case of a flat extra dimension where all the mode expansions are ordinary Fourier series and the corresponding boundary conditions are simply Neumann or Dirichlet.} as they must be if they are the Goldstone modes of the broken Kac-Moody algebra. As emphasized in \cite{Lim:2005rc,Lim:2007fy,Lim:2008hi}, the $N=2$ SUSY structure relating the modes also determines the boundary conditions of these modes -- and we generalize that construction here to uncover the SUSY structure and diffeomorphism invariance of the RS model with brane-localized curvature terms.

We begin by writing the KK decomposition of the metric fluctuations as \cite{Lim:2005rc,Lim:2007fy,Lim:2008hi,Chivukula:2022kju,Chivukula:2023qrt}
\begin{eqnarray}
    h_{\mu\nu}(x^\alpha,z) =&& \sum\limits_{n=0}^{\infty}h_{\mu\nu}^{(n)}(x^\alpha)f^{(n)}(z),\label{eq:KK_1u}\\
    A_{\mu}(x^\alpha,z) =&& \sum\limits_{n=1}^{\infty}A_{\mu}^{(n)}(x^\alpha)g^{(n)}(z),\label{eq:KK_2u}\\
    \varphi(x^\alpha,z) =&&~ r(x^\alpha)k^{(0)}(z) +  \sum\limits_{n=1}^{\infty}\pi^{(n)}(x)k^{(n)}(z)~,\label{eq:KK_3u}
\end{eqnarray}
where the fields $h^{(n)}_{\mu\nu}$ and $r$ are the physical four-dimensional spin-2 KK fields and the radion, while the $A^{(n)}_\mu$ and $\pi^{(n)}$ are the corresponding vector and scalar Goldstone fields respectively.
Since the brane-localized curvature terms do not change the equation of motion in the bulk, 
the form of the Sturm-Liouville problems associated with these modes will have the same form as in RS1 and can be written in the form of the quantum-mechanical SUSY relations
\begin{equation}
    \begin{cases}
        D f^{(n)} = m_ng^{(n)}~, \\
        D^\dagger g^{(n)} = m_nf^{(n)}~,
    \end{cases}
    \qquad
    \begin{cases}
        \overline{D} g^{(n)} = m_nk^{(n)} \\
        \overline{D}^\dagger k^{(n)} = m_ng^{(n)}~.
    \end{cases}
    \label{eq:SUSY-SLsystem}
\end{equation}
Here the operators $D$ and $\bar{D}$ are defined as in Eq.~(\ref{eq:D-definitions}) and the masses $m_n$ are the masses of the physical spin-2 fields.\footnote{For simplicity, we restrict our attention here to an unstabilized RS model with a massless radion. The generalization of the SUSY analysis to a model with brane-localized curvature terms {\it and} stabilized by the Goldberger-Wise \cite{Goldberger:1999uk,Goldberger:1999un} mechanism could be constructed in analogy to the analysis given in \cite{Chivukula:2022kju}.} 

However, the boundary conditions satisfied by the mode wavefunctions are modified by the brane-localized curvature terms. As emphasized by \cite{Lim:2005rc,Lim:2007fy,Lim:2008hi}, the boundary conditions are determined by insuring that there are suitable Hermitian conjugate supercharges built, as we will see, from the operators in Eq.~(\ref{eq:D-definitions}), with respect to the inner product needed to make the kinetic energy terms of the fields canonical. From Eqs.~(\ref{eq:RS1hh}), (\ref{eq:RS1AA}) and (\ref{eq:RS1phiphi}), we see that, in order to have orthogonal KK modes, the mode function inner-products now should take the forms,
\begin{eqnarray}
    \delta_{m,n}=\braket{f^{(m)} | f^{(n)}}_f &\equiv& \int_{z_1}^{z_2}dz~e^{3A(z)} f^{(m)}(z)f^{(n)}(z)\left(1+\sum_i\gamma_i\delta_i\right),
    \label{eq:h-inner-product}\\
    \delta_{m,n}=\braket{g^{(m)} | g^{(n)}}_g &\equiv& \int_{z_1}^{z_2}dz~e^{3A(z)} g^{(m)}(z)g^{(n)}(z),\\
    \delta_{m,n}=\braket{k^{(m)} | k^{(n)}}_k &\equiv& \int_{z_1}^{z_2}dz~e^{3A(z)} k^{(m)}(z)k^{(n)}(z)\left(1+\sum_i\dfrac{\gamma_i\delta_i}{1-2A'\gamma_i\theta_i}\right).
    \label{eq:k-inner-product}
\end{eqnarray}

Note also that, in order for inner products to be positive-definite and thus for the Sturm-Liouville problem to be regular, the weight function must be positive definite on the interval, and hence
\begin{equation}
    \gamma_i \ge 0,\qquad \dfrac{\gamma_i}{1-2A'\gamma_i\theta_i} \ge 0.
    \label{eq:gamma-bound}
\end{equation}
If these conditions were not satisfied there would exist states with negative norm, violating unitarity. Therefore, the parameters for the brane-localized curvature terms must satisfy
\begin{equation}
    \gamma_1 \ge 0,\qquad 0\le\gamma_2<\dfrac{z_2}{2}.
\end{equation}
The upper bound in the second of these conditions, which arises from positivity of the weight function in the scalar sector, is equivalent to that found in \cite{Davoudiasl:2003zt,Luty:2003vm,George:2011sw}. The requirement that both coefficients must satisfy $\gamma_i\ge 0$, which arises from the positivity of the weight function in the spin-2 sector, was recently independently reported in \cite{Miao:2023mui}.

To derive the SUSY-compatible boundary conditions \cite{Lim:2005rc,Lim:2007fy,Lim:2008hi,Chivukula:2022kju}, we consider a SUSY doublet 
\begin{equation}
    \Psi = \begin{pmatrix} f \\ g \end{pmatrix},
\end{equation}
with the inner product defined as
\begin{equation}
\aligned
    \braket{\tilde{\Psi} | \Psi} =&~ \int_{z_1}^{z_2}dz~e^{3A} \tilde{f}f \left(1+\sum_i\gamma_i\delta_i\right)  + \int_{z_1}^{z_2}dz~e^{3A} \tilde{g}g.
\endaligned
\end{equation}
The supercharges are defined as 
\begin{equation}
    Q = \begin{pmatrix} 0&0\\D&0 \end{pmatrix},\qquad Q^\dagger = \begin{pmatrix} 0&D^\dagger\\0&0 \end{pmatrix}.
\end{equation}
In order for the boundary conditions to respect the supersymmetry as well, the supercharges $Q$ and $Q^\dagger$ must be the hermitian conjugates of each other with respect to the inner product,
\begin{equation}
    \braket{\tilde{\Psi}|Q\Psi} = \braket{Q^\dagger\tilde{\Psi}|\Psi}.
\end{equation}
Thus, one can derive the boundary conditions
\begin{equation}
    g(z_i)+\theta_i\gamma_iD^\dagger g(z_i) = 0,\quad {\rm and}\quad Df(z_i)+\theta_i\gamma_iD^\dagger Df(z_i) = 0,\label{eq:bc_1}
\end{equation}
or
\begin{equation}
    f(z_i) = 0,\quad {\rm and}\quad D^\dagger g(z_i)=0.
\end{equation}
Similarly, for the $g$-$k$ SUSY doublet, one can derive the following SUSY-compatible boundary conditions,
\begin{equation}
    g(z_i)-\dfrac{\gamma_i}{\theta_i-2A'\gamma_i}\overline{D} g(z_i) = 0,\quad {\rm and}\quad \overline{D}^\dagger k(z_i)-\dfrac{\gamma_i}{\theta_i-2A'\gamma_i}\overline{D} \overline{D}^\dagger k(z_i) = 0, \label{eq:bc_2}
\end{equation}
or
\begin{equation}
    k(z_i) = 0,\quad {\rm and}\quad \overline{D} g(z_i)=0.
\end{equation}

Notice that the boundary conditions for $g$ in Eqs.~(\ref{eq:bc_1}) and (\ref{eq:bc_2}) are actually the same,
\begin{equation}
    g(z_i)+\theta_i\gamma_iD^\dagger g(z_i) = \left(\theta_i-2A'\gamma_i\right)\left[g(z_i)-\dfrac{\gamma_i}{\theta_i-2A'\gamma_i}\overline{D} g(z_i)\right].
\end{equation}
Any other choice will not allow for the  $N=2$ SUSY relations for {\it both} the $f$-$g$ and $g$-$k$ systems to hold. Therefore, the unique set of boundary conditions for the mode expansions of the RS model with brane-localized curvature terms are
\begin{equation}
    \begin{cases}
        Df(z_i)+\theta_i\gamma_iD^\dagger Df(z_i) = 0\\
        g(z_i)+\theta_i\gamma_iD^\dagger g(z_i) = 0\\
        \overline{D}^\dagger k(z_i)-\dfrac{\gamma_i}{\theta_i-2A'\gamma_i}\overline{D} \overline{D}^\dagger k(z_i) = 0
    \end{cases}.
    \label{eq:bc_fgk}
\end{equation}
In the limit $\gamma_{1,2} \to 0$, these conditions reduce to those previously found by imposing the $N=2$ SUSY conditions without brane-localized curvature terms \cite{Lim:2005rc,Lim:2007fy,Lim:2008hi,Chivukula:2022kju}.

Combining the SUSY relations in Eq.~(\ref{eq:SUSY-SLsystem}) with these boundary conditions we can derive the Sturm-Liouville eigenmode systems associated with the spin-2, spin-1, and spin-0 fields. For the $f^{(n)}(z)$ functions of the spin-2 system, for example, we find the Sturm-Liouville systems
\begin{align}
    D^\dagger D f^{(n)}(z)  = m_n D^\dagger(g^{(n)}(z)) & =m^2_n f^{(n)}(z) \label{eq:spin2-eq1}\\
    Df^{(n)}(z_{1,2})+ m^2_n \theta_i\gamma_i f^{(n)}(z_{1,2}) & = 0~, \label{eq:spin2-eq2}
\end{align}
where we have used the eigenvalue equation to write the boundary-conditions for each mode in a conventional form. We see that this spin-2 Sturm-Liouville problem has the weight implied by the inner-product in Eq.~(\ref{eq:h-inner-product}), and reproduces the spin-2 (unitary gauge) mode equations derived previously in the literature \cite{Davoudiasl:2003zt,Shtanov:2003um,Carena:2005gq,Bao:2005bv,George:2011sw,Dillon:2016bsb}. As we describe in the following section, the analysis given here allows us to derive the Sturm-Liouville systems associated with the spin-1 and spin-0 modes which maintain consistency with the underlying gravitational diffeomorphism symmetries \cite{Lim:2005rc,Lim:2007fy,Lim:2008hi,Chivukula:2022kju}.

For the zero modes, the SUSY relations become
\begin{equation}
    Df^{(0)}(z) = D^\dagger g^{(0)}(z) = \overline{D}^\dagger k^{(0)}(z) = 0.
\end{equation}
Thus, their boundary conditions are not affected by the brane-localized curvature terms, and their wavefunctions are same as the ones in RS1,
\begin{equation}
    f^{(0)}(z) = {\rm Const},\quad k^{(0)}(z) = \mathcal{N}e^{-2A(z)},\quad g^{(0)}(z) = 0~.
    \label{eq:zero-mode-wavefunctions}
\end{equation}

%%%%%%%%%%%%%%%%%%%%%%%%%%%%%%%%%%%%%%%%%%%%%%%%%%%%%%
\section{5D Diffeomorphism Invariance of the Kaluza-Klein Theory}

\label{sec:5Ddiffeomorphism}

In this section we show that the  incorporation of the auxiliary field $\Delta(x^\mu,z)$, subject to the constraint in Eq.~(\ref{eq:P3}), reconciles the boundary conditions on the metric fluctuation mode expansions determined above from their $N=2$ SUSY structure with the diffeomorphism invariance of the RS1 Kaluza-Klein theory background geometry.

 Under an infinitesimal coordinate transformation, Eq.~(\ref{eq:linear-diffeomorphisms}), the metric transforms as 
 \begin{align}
     G_{M N} \mapsto G_{M N}-G_{M A} \partial_N \xi^A-G_{N A} \partial_M \xi^A-\xi^A \partial_A G_{M N}.
     \label{eq:metric-transformation}
 \end{align}
 Adding brane-localized curvature terms does not change the background geometry, therefore the diffeomorphism invariances of the model are identical to those in the RS1 model without brane-localized curvature terms \cite{Lim:2005rc,Lim:2007fy,Lim:2008hi,Chivukula:2022kju}, and satisfy the boundary conditions
\begin{equation}
    \partial_z \xi_\mu(z_i) = \xi^5(z_i) = 0,
    \label{eq:xi-boundary-conditions}
\end{equation}
that is, Neumann boundary conditions for $\xi_\mu$ and Dirichlet boundary conditions for $\xi^5$. 
These transformations respect the ``straight gauge" \cite{Carena:2005gq} conditions, since the locations of the boundaries are fixed and the conditions of Eq.~(\ref{eq:straight-gauge}) are satisfied. 

In the absence of brane-localized curvature terms (the limit $\gamma_{1,2}\to 0$ in Eq.~(\ref{eq:bc_fgk})), the boundary conditions on the $(f^{(n)}, g^{(n)})$  system are also Neumann and Dirichlet, respectively. In this case we can expand the infinitesimal diffeomorphism parameters $(\xi_\mu, \xi^5)$ in the same mode expansion as we use for the metric fluctuation fields. Consequently the metric transformations in Eq.~(\ref{eq:metric-transformation}) act simply on the metric fluctuation fields -- allowing one, for example, to easily identify the spin-1 and spin-0 fields as unphysical Goldstone bosons and construct Feynman-like gauge-fixing terms \cite{Lim:2005rc,Lim:2007fy,Lim:2008hi,Chivukula:2022kju}.

However, in the presence of brane-localized curvature terms, the $(f^{(n)}, g^{(n)})$ now satisfy the boundary conditions given in Eq.~(\ref{eq:bc_fgk}).
The $(\xi_\mu,\xi^5)$ and the $(f^{(n)}, g^{(n)})$ therefore satisfy  different boundary conditions and we cannot expand the allowed infinitesimal coordinate transformations $(\xi_\mu,\xi^5)$ using the mode eigenfunctions $(f^{(n)}, g^{(n)})$. The relationship between the diffeomorphism transformations and the transformation properties of the fields are therefore modified, as we show below.

At the linearized level, the field transformation in Eq.~(\ref{eq:metric-transformation}) can be written as,
\begin{eqnarray}
     \delta h_{\mu\nu}+\dfrac{1}{\sqrt{6}}\partial_\mu\partial_\nu \delta \Delta+\dfrac{\eta_{\mu\nu}}{2\sqrt{6}}\left(\delta \Delta''+3A'\delta \Delta'\right)&= & - \partial_\mu\xi_\nu - \partial_\nu\xi_\mu - \eta_{\mu\nu}(\partial_z+3A')\xi^5,\label{eq:diffeomorphism-begin}\\
     \epsilon\,\delta A_{\mu}&= & - \sqrt{2}\partial_z\xi_\mu + \sqrt{2}\partial_\mu\xi^5,\\
     \delta \varphi + \dfrac{1}{2}\left(\delta \Delta''+A'\delta \Delta'\right)&= & - \sqrt{6}(\partial_z+A')\xi^5 \label{eq:diffeomorphism-end},
\end{eqnarray}
where we now allow for the auxiliary field $\Delta$ to vary under a diffeomorphism transformation.
While the $\xi_\mu$ and $\xi^5$ on the right hand of the above equations satisfy the Neumann and Dirichlet boundary conditions, the gravitational field $\delta h_{\mu\nu}$, $\delta A_\mu$ and $\delta \varphi$ on the left hand obey the  boundary conditions given in Eqs.~(\ref{eq:bc_fgk}). The auxiliary field $\Delta$ must transform in a way that makes these compatible. 

As we now show, the needed transformation of $\Delta$ preserves the condition in Eq.~(\ref{eq:P3}) -- demonstrating the diffeomorphism invariance of the Kaluza-Klein theory in the presence of brane-curvature terms. 
Conveniently, we can rewrite the above transformation as
\begin{eqnarray}
     \delta h_{\mu\nu}&= & - \partial_\mu\tilde{\xi}_\nu - \partial_\nu\tilde{\xi}_\mu - \eta_{\mu\nu}(\partial_z+3A')\tilde{\xi}^5,\label{eq:h-transformation}\\
     \delta A_{\mu}&= & - \sqrt{2}\partial_z\tilde{\xi}_\mu + \sqrt{2}\partial_\mu\tilde{\xi}^5,\\
     \delta \varphi&= & - \sqrt{6}(\partial_z+A')\tilde{\xi}^5, \label{eq:phi-transformation}
\end{eqnarray}
where the new parameters $\tilde{\xi}$ are defined as
\begin{equation}
    \tilde{\xi}_\mu \equiv \xi_\mu + \dfrac{1}{2\sqrt{6}}\partial_\mu\delta \Delta, \qquad \tilde{\xi}^5 \equiv \xi^5 + \dfrac{1}{2\sqrt{6}}\delta \Delta'.
    \label{eq:tilde-xi}
\end{equation}
The auxiliary field $\Delta$ must transform in a way such that $\tilde{\xi}_\mu$ satisfies the boundary condition for $f^{(n)}$, and $\tilde{\xi}^5$ satisfies the boundary condition for $g^{(n)}$,
\begin{eqnarray}
    D\tilde{\xi}_\mu+\theta_i\gamma_iD^\dagger D\tilde{\xi}_\mu = 0,\\
    \tilde{\xi}^5+\theta_i\gamma_iD^\dagger \tilde{\xi}^5. = 0. \label{eq:bc_xi5}
\end{eqnarray}
Combining Eq.~(\ref{eq:bc_xi5}) and the $\xi^5$ condition in Eq.~(\ref{eq:xi-boundary-conditions}), one can derive the following constraint on $\delta \Delta$,
\begin{equation}
    \delta \Delta'(z_i) = \dfrac{2\sqrt{6}\gamma_i}{\theta_i-2A'\gamma_i}(\partial_z+A')\left(\xi^5+\dfrac{1}{2\sqrt{6}}\delta \Delta'\right) = -\dfrac{2\gamma_i}{\theta_i-2A'\gamma_i}\delta \varphi.
    \label{eq:dP3}
\end{equation}
Therefore, as promised, Eq.~(\ref{eq:P3}), which is necessary for diagonalizing the quadratic Lagrangian, is covariant under the residual symmetry transformation. For an arbitrary infinitesimal diffeomorphism transformation $(\xi_\mu,\xi^5)$ consistent with straight gauge, therefore, the metric fluctuation field transformations are given by the Eqs.~(\ref{eq:h-transformation}) - (\ref{eq:phi-transformation}) in terms of $(\tilde{\xi}_\mu, \tilde{\xi}^5)$ as defined in Eq.~(\ref{eq:tilde-xi}). 

Furthermore, $\delta \Delta$ is unconstrained in the bulk, and we are free to choose any function for $\delta \Delta$, as long as it satisfies the constraint on the boundaries in Eq.~(\ref{eq:dP3}). As a consequence, one can easily see that $\Delta$ is also unconstrained in the bulk as we argued in the previous section, since any two different choices of $\Delta$, denoted as $\Delta_{a}$ and $\Delta_{b}$ are related by the diffeomorphism transformation,
\begin{equation}
    \xi_\mu = -\dfrac{1}{2\sqrt{6}} \delta \partial_\mu \Delta = -\dfrac{1}{2\sqrt{6}} \partial_\mu(\Delta_{a}-\Delta_{b}),\quad \xi^5 = -\dfrac{1}{2\sqrt{6}} \delta \Delta' = -\dfrac{1}{2\sqrt{6}} (\Delta_{a}'-\Delta_{b}').
\end{equation}

From the form of Eq.~(\ref{eq:tilde-xi}) and the fact that $\Delta$ is arbitrary in the bulk we see that the auxilary field can be viewed as an constrained, $\varphi$-dependent,  ``brane-bending" mode necessary to appropriately diagonalize the scalar metric fluctuations in the presence of brane-curvature interactions, as shown in Eqs.~(\ref{eq:mode-mixing})-(\ref{eq:P3}).

\section{Gauge-Fixing}
\label{sec:GF}

Once we expand $\tilde{\xi}_\mu$ and $\tilde{\xi}^5$ using the corresponding mode eigenfunctions,
\begin{eqnarray}
     \tilde{\xi}_\mu(x^\alpha,z) &=& \sum\limits_{n=0}^{\infty}\tilde{\xi}^{(n)}_\mu(x^\alpha)f^{(n)}(z),\\
     \tilde{\xi}^5(x^\alpha,z) &= &\sum\limits_{n=0}^{\infty}\tilde{\xi}^{5(n)}(x^\alpha)g^{(n)}(z)~,
     \label{eq:diffeomorphism-modes}
\end{eqnarray}
the transformations on the individual KK modes defined in Eqs.~(\ref{eq:KK_1u}) - (\ref{eq:KK_3u}) can be written as 
\begin{eqnarray}
     h_{\mu\nu}^{(n)}&\mapsto & h_{\mu\nu}^{(n)} - \partial_\mu\tilde{\xi}_\nu^{(n)} - \partial_\nu\tilde{\xi}_\mu^{(n)} + m_n \eta_{\mu\nu}\tilde{\xi}^{5(n)},\label{eq:diffeomorphism-spin2}\\
     A_{\mu}^{(n)}&\mapsto & A_{\mu}^{(n)} - \sqrt{2}m_n\tilde{\xi}_\mu^{(n)} + \partial_\mu\tilde{\xi}^{5(n)},\label{eq:diffeomorphism-spin1}\\
     \pi^{(n)}&\mapsto & \pi^{(n)} - \sqrt{6}m_n\tilde{\xi}^{5(n)},\\
     r&\mapsto & r.
\end{eqnarray}
These transformations are identical to those found in RS1 without brane-localized curvature terms \cite{Lim:2005rc,Lim:2007fy,Lim:2008hi,Chivukula:2022kju}, and gauge-fixing can proceed analogously.

Unitary gauge can be achieved by choosing
\begin{alignat}{2}
     \tilde{\xi}^{(n)}_\mu &=~ \dfrac{1}{\sqrt{2}m_n}\left(A_\mu^{(n)} + \dfrac{1}{\sqrt{6}m_n}\partial_\mu\pi^{(n)}\right),
     \quad &n\geq1,\label{eq:unitary1}\\
     \tilde{\xi}^{5(n)} &=~ \dfrac{1}{\sqrt{6}m_n}\pi^{(n)},\quad &n\geq 1.\label{eq:unitary2}
\end{alignat}
Alternatively, the gauge redundancy can be removed by introducing the 5D $R_\xi$ gauge fixing term
\begin{equation}
    \mathcal{L}_{\rm GF} = \dfrac{e^{3A}}{\xi}\left[F_\mu F^\mu\left(1+\sum_i\gamma_i\delta_i\right) - F_5 F_5\right],
    \label{eq:tHooft-gauge}
\end{equation}
where
\begin{eqnarray}
    F_\mu &=& -\left[\partial^\nu h_{\mu\nu} - \dfrac{1}{2}\left(2-\dfrac{1}{\xi}\right)\partial_\mu h^\nu_\nu + \dfrac{\xi}{\sqrt{2}}D^\dagger A_\mu\right], \label{eq:Fmu} \\
    F_5 &=& -\left(\dfrac{1}{2}D h^{\mu}_\mu - \dfrac{1}{\sqrt{2}}\partial_\mu A^\mu +\xi\sqrt{\dfrac{3}{2}}\,\overline{D}^\dagger\varphi\right).\label{eq:F5}
\end{eqnarray}
These gauge-fixing terms are a straightforward generalization of the gauge-fixing given in \cite{Lim:2005rc,Lim:2008hi,Chivukula:2022kju}, generalized to account for the modified weight function of the $f^{(n)}$ inner product shown in Eq.~(\ref{eq:h-inner-product}). 

To demonstrate that physical results are $\xi$-independent, we consider the example of  tree-level correlation functions between
two conserved energy-momentum tensors $T^{(i)\mu\nu}$ residing on the two branes at $z=z_i$, the correlation functions which determine the gravitational interactions of brane-localized matter.\footnote{This problem was examined in \cite{Lim:2008hi} in the absence of brane-localized curvature terms, and we generalize their computation here.} The interaction between the gravitational fields and the external currents in this model are given by
\begin{equation}
    -\dfrac{\kappa}{2}\sum_i \int d^4x\,H_{\mu\nu}(z_i) T^{(i)\mu\nu} = -\dfrac{\kappa}{2} \sum_i \int d^4x\left[h_{\mu\nu} - \dfrac{1}{\sqrt{6}}\eta_{\mu\nu}\left(\varphi -A' \Delta'\right) +\dfrac{1}{\sqrt{6}}\partial_\mu\partial_\nu \Delta\right]_{z=z_i} T^{(i)\mu\nu}.
    \label{eq:tmunu-coupling}
\end{equation}
Note that one must properly include the couplings of the auxiliary field $\Delta$, however the last term in the bracket does not contribute since $T^{(i)\mu\nu}$ is a conserved current. Furthermore, the remaining $\Delta$ interactions only depend on the values of $\Delta'(z_{1,2})$, so we can immediately replace these with a coupling to the field $\varphi$ using Eq~(\ref{eq:P3}): we do not have to specify the value of $\Delta$ in the bulk for this calculation. The final form of the gravitational interactions with the external currents are hence
\begin{equation}
    -\dfrac{\kappa}{2}\sum_i \int d^4x\left[h_{\mu\nu} - \dfrac{1}{\sqrt{6}}\eta_{\mu\nu}\dfrac{\theta_i}{\theta_i-2A'\gamma_i}\varphi\right]_{z=z_i} T^{(i)\mu\nu}.
\end{equation}
Following Ref.~\cite{Lim:2008hi}, the gauge dependent part of the amplitude can be written as
\begin{equation}
    \mathcal{M}\supset -\sum_{n>0}\dfrac{\kappa^2}{24}\dfrac{\tilde{T}^{(1)}(p)\tilde{T}^{(2)}(-p)}{p^2-(3\xi-2)m_n^2}\left[-f^{(n)}(z_1)f^{(n)}(z_2)+\dfrac{\theta_1}{\theta_1-2A'\gamma_1}k^{(n)}(z_1)\dfrac{\theta_2}{\theta_2-2A'\gamma_2}k^{(n)}(z_2)\right],
    \label{eq:xi-dependent}
\end{equation}
where $\tilde{T}^{(i)}$ is the Fourier transform of the trace of the energy momentum tensor. Note that this contribution has spurious $\xi$-dependent poles which must cancel out for any physical amplitude. 

Using the $N=2$ SUSY relations in Eq.~(\ref{eq:SUSY-SLsystem}) and the definitions of the differential operators in Eq.~(\ref{eq:D-definitions}), one can write the scalar wavefunction $k^{(n)} (n>0)$ in terms of $f^{(n)}$ and $g^{(n)}$,
\begin{equation}
\aligned
    k^{(n>0)}(z_i) =&\dfrac{\bar{D}}{m_n}g^{(n)}(z_i)= \dfrac{(-D^\dagger - A')}{m_n} g^{(n)}(z_i)= - f^{(n)}(z_i) - \dfrac{2A'}{m_n}g^{(n)}(z_i)\\
    =& -\dfrac{\theta_i-2A'\gamma_i}{\theta_i} f^{(n)}(z_i) -\dfrac{2A'}{m_n}\left(g^{(n)}(z_i)+\theta_i\gamma_i m_n f^{(n)}(z_i)\right),
\endaligned
\end{equation}
where the last term vanishes on the branes due to the boundary condition in Eq.~(\ref{eq:bc_1}). Therefore, the $\xi$-dependent part of the amplitude shown in Eq.~(\ref{eq:xi-dependent}) vanishes as a result of the cancellation between the spin-2 and spin-0 contributions,
\begin{equation}
    \left[-f^{(n)}(z_1)f^{(n)}(z_2)+\dfrac{\theta_1}{\theta_1-2A'\gamma_1}k^{(n)}(z_1)\dfrac{\theta_2}{\theta_2-2A'\gamma_2}k^{(n)}(z_2)\right]=0.
\end{equation}

This proof of the gauge-invariance of the energy-momentum tensor correlation function would fail if we considered matter localized at an arbitrary point in the bulk, as we now show. Since the value of $\Delta$ is arbitrary in the bulk, we can choose it (and its derivatives) to vanish at the location $z\neq z_i$ where the external field is localized.\footnote{In fact, as we show in the next section, there is a convenient choice of the auxiliary field $\Delta$ as a distribution which vanishes everywhere in the bulk - see Eq.~(\ref{eq:P3_delta}).} In this case the couplings of the metric to an external source in the bulk (of the form of Eq.~(\ref{eq:tmunu-coupling}), but localized to a point in the bulk) yields only a coupling to $\varphi$.
The coupling of the external field would then be  proportional to
\begin{align}
     k^{(n>0)}(z) =& - f^{(n)}(z) - \dfrac{2A'}{m_n}g^{(n)}(z)~,
\end{align}
and the gauge-dependent parts of the correlation function in Eq.~(\ref{eq:xi-dependent}) will in general be proportional to $A'(z)g^{(n)}(z) \neq 0$. In general, therefore, one cannot couple five-dimensional gravity to matter confined to a brane at an {\it arbitrary location} in the bulk and maintain five-dimensional diffeomorphism invariance. This result is consistent with the finding in \cite{Chivukula:2023sua} that couplings of gravity to matter localized at an arbitrary point in the bulk have anomalous ${\cal O}(s^3)$ scattering amplitudes with KK gravitons.

In 't-Hooft-Feynman gauge, choosing $\xi=1$, the gauge-fixed kinetic terms are given by
\begin{equation}
\aligned
    S = \int d^4x~\sum_n\left\{\vphantom{\dfrac{1}{2}}\right.&\dfrac{1}{2}h^{(n)}_{\mu\nu}\left[\dfrac{1}{2}\left(\eta^{\mu\rho}\eta^{\nu\sigma} + \eta^{\mu\sigma}\eta^{\nu\rho} - \eta^{\mu\nu}\eta^{\rho\sigma}\right)(-\Box - m_n^2)\right]h^{(n)}_{\rho\sigma} \\
    & + \dfrac{1}{2}A^{(n)}_{\mu}\left[-\eta^{\mu\nu}(-\Box - m_n^2)\right]A^{(n)}_{\nu} \\
    & + \dfrac{1}{2}~\pi^{(n)}\left(-\Box - m_n^2\right)\pi^{(n)}\left.\vphantom{\dfrac{1}{2}}\right\} + \dfrac{1}{2}~r\left(-\Box \right)r.
\endaligned
\label{eq:gauge-fixed-ke}
\end{equation}
Note that as a consequence of the SUSY conditions all of the particles at each level $n$ are degenerate and there are no ``spurious" poles at unphysical masses.

\section{Scattering amplitudes}

\label{sec:Scattering}

In this section, we study the high-energy behavior of the elastic scattering amplitude of longitudinally polarized (helicity-0) KK gravitons,
\begin{equation}
    h_L^{(n)}h_L^{(n)}\rightarrow h_L^{(n)} h_L^{(n)},
    \label{eq:four-longitudinal}
\end{equation}
generalizing the analyses of \cite{SekharChivukula:2019yul,SekharChivukula:2019qih,Chivukula:2020hvi} to the model with brane-localized curvature terms.
In the following subsection we report the results of a direct computation of this amplitude in unitary gauge, and find that it grows like ${\cal O}(s^3)$. In the subsequent subsection we show that the the spin-0 Goldstone bosons corresponding to the helicity-0 states of the massive spin-2 KK bosons behave as a tower of Galileons \cite{Dvali:2000hr,Luty:2003vm,Nicolis:2008in,deRham:2010ik,Hinterbichler:2011tt,deRham:2014zqa}, explaining (via the Goldstone boson equivalence theorem) the high-energy behavior of the amplitude. In the last subsection we investigate how the scale $\Lambda_3$, characterizing the Goldstone boson Galileon interactions, relates to the other mass scales of the theory.

\subsection{Scattering in Unitary Gauge and ${\cal O}(s^5,s^4)$ Sum Rules \label{subsec:unitary}}

Using the formalism developed above we have performed the analytic and numerical computations of the scattering amplitudes for Eq.~(\ref{eq:four-longitudinal}) in unitary gauge. Specifically, in this gauge the only physical particles are the massive spin-2 KK tower and a massless radion. The spin-2 mass spectrum and mode functions are determined by the Sturm-Liouville system of Eqs.~(\ref{eq:spin2-eq1}) and (\ref{eq:spin2-eq2}) normalized via Eq.~(\ref{eq:h-inner-product}), the wavefunction of the radion is given by $k^{(0)}(z)$ shown in Eq.~(\ref{eq:zero-mode-wavefunctions}) subject to the normalization condition in Eq.~(\ref{eq:k-inner-product}), and $\Delta(z)$ must satisfy the boundary conditions of Eq.~(\ref{eq:P3}) but is otherwise arbitrary. Following the analyses in \cite{SekharChivukula:2019yul,SekharChivukula:2019qih,Chivukula:2020hvi} we analyze the energy dependence of the scattering amplitude by expanding the helicity-0 matrix element $\mathcal{M}$ as at large energies in terms of the scattering energy $\sqrt{s}$ and the scattering angle $\theta$,
\begin{equation}
    \mathcal{M}(s,\theta) = \sum_{\sigma\in \mathbb{Z}}\widetilde{\mathcal{M}}^{(\sigma)}(\theta) s^{\sigma/2},
\end{equation}
and examine the ``reduced" matrix elements $\widetilde{\mathcal{M}}^{(\sigma)}$.

At the order of $\mathcal{O}(s^5)$, the reduced amplitude can be written as 
\begin{equation}
\aligned
    \widetilde{\mathcal{M}}^{(10)} =~& \frac{\kappa^2 (\cos2\theta+7)\sin^2\theta}{2304 m_n^8}\left[\sum_{j=0}^{\infty}a_{nnj}^2 - a_{nnnn}\right],
    \label{eq:s5}
\endaligned
\end{equation}
where the couplings are defined as
\begin{eqnarray}
    a_{nnj} &=& \braket{f^{(n)}f^{(n)}f^{(j)}}_f\equiv\int_{z_1}^{z_2}dz~e^{3A}f^{(n)}(z)f^{(n)}(z)f^{(j)}(z)\left(1+\sum_{i=1}^2\gamma_i\delta_i\right), \\
    a_{nnnn} &=& \braket{f^{(n)}f^{(n)}f^{(n)}f^{(n)}}_f\equiv \int_{z_1}^{z_2}dz~e^{3A}f^{(n)}(z)f^{(n)}(z)f^{(n)}(z)f^{(n)}(z)\left(1+\sum_{i=1}^2\gamma_i\delta_i\right).
\end{eqnarray}
The form of the expression for $\widetilde{\mathcal{M}}^{(10)}$ agrees with that found in the RS model without brane-localized curvature terms \cite{SekharChivukula:2019yul,SekharChivukula:2019qih,Chivukula:2020hvi}, with the couplings generalized to account for the revised inner product of Eq.~(\ref{eq:h-inner-product}).

The coupling constant combination appearing in Eq.~(\ref{eq:s5}) vanishes precisely as in the model with brane-localized curvature \cite{SekharChivukula:2019qih,Bonifacio:2019ioc}.
In particular, the wavefunctions $\{f^{(j)}\}$ form a complete basis with respect to the weight function $e^{3A}(1+\sum_{i=1}^2\gamma_i\delta_i)$,
\begin{equation}
    \left(f^{(n)}(z')\right)^2 = \sum_{j=0}^{\infty}f^{(j)}(z') \int_{z_1}^{z_2}dz~e^{3A}\left(f^{(n)}(z)\right)^2f^{(j)}(z)\left(1+\sum_{i=1}^2\gamma_i\delta_i\right), \quad {\rm for}~z_1<z'<z_2.
\end{equation}
This eigenfunction expansion is strictly only valid for $z_1<z'<z_2$ but not on the boundary $z'=z_{1,2}$, since $\left(f^{(n)}(z)\right)^2$ does not satisfy the boundary condition that $f^{(j)}(z)$ does. However, one can still use the generalized Parseval's identity to show that it converges upon integration
\begin{equation}
    \sum_{j=0}^{\infty}a_{nnj}^2 - a_{nnnn} = 0.
\end{equation}
Similarly, at the order of $\mathcal{O}(s^4)$, the reduced amplitude can be written as 
\begin{equation}
\aligned
    \widetilde{\mathcal{M}}^{(8)} =~& \frac{\kappa^2 (\cos2\theta+7)}{9216 m_n^6}\left[\dfrac{4}{3}a_{nnnn} - \sum_{j=0}^{\infty}\dfrac{m_j^2}{m_n^2}a_{nnj}^2 \right].
\endaligned
\end{equation}
This expression also agrees with the RS model without brane-localized curvature terms, and 
also vanishes once the mode equations and the completeness are applied \cite{SekharChivukula:2019qih,Bonifacio:2019ioc}.

While the presence of the brane-localized curvature terms changes the details of spin-2 KK masses and wavefunctions through their effect on the Sturm-Liouville system in Eqs.~(\ref{eq:spin2-eq1}) and (\ref{eq:spin2-eq2}), these computations show that the extra scalar interactions induced by the appearance of auxiliary field $\Delta(x^\mu,z)$ in the metric (Eq.~(\ref{eq:background-metric})) and its dependence on the field $\varphi(x^\mu,z)$ (from Eq.~(\ref{eq:P3})) {\it do not} change the fact that the amplitudes vanish to  ${\cal O}(s^5,s^4)$

However, unlike in RS1 without brane-curvature terms, the reduced amplitude at the order of $\mathcal{O}(s^3)$, does not vanish,
    $\widetilde{\mathcal{M}}^{(6)} \ne~ 0$.
We have verified this both numerically and analytically, and checked that the value of the scattering amplitude is independent of the value chosen for the auxiliary field $\Delta$ in the bulk. The form of this reduced amplitude can be more easily computed and displayed in a compact manner in `t-Hooft-Feynman gauge, as we show in the next subsection.

\begin{figure}[tbh]
   % \centering
    \includegraphics[width=0.48\textwidth]{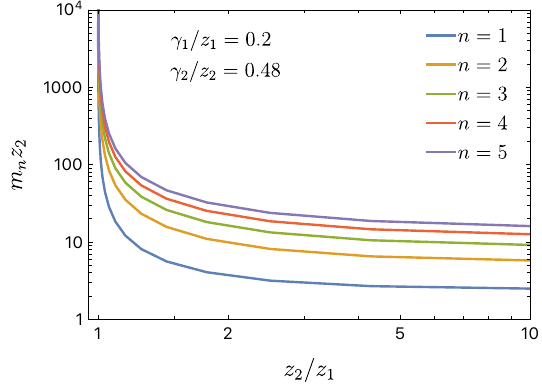}
    \includegraphics[width=0.48\textwidth]{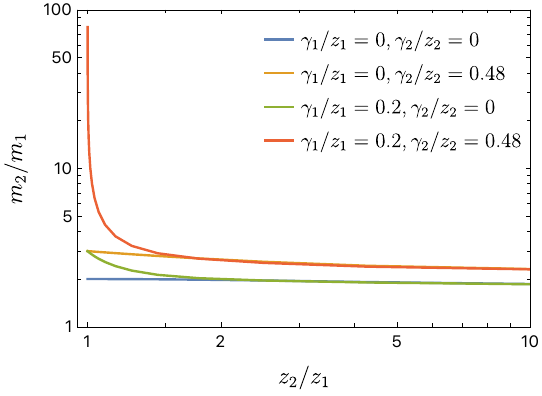}
    \caption{Left: The masses of the first few spin-2 KK states for different AdS ``hierarchies" (ratios $z_2/z_1$) and brane parameters $\gamma_i$ bounded by Eq.~(\ref{eq:gamma-bound}) Right: The ratio $m_2/m_1$ as a function of $z_2/z_1$ for values of the brane-localized curvature interaction. Note that the masses are more sensitive to the value of the IR brane-localized curvature strength $\gamma_2$ \cite{Davoudiasl:2003zt} than that of the UV brane-localized curvature term, but the ratio of the first two spin-2 KK masses can exceed 2 for $\gamma_2$ close to maximal or if the AdS hierarchy is relatively small.}
    \label{fig:masses}
\end{figure}

We have also checked explicitly that the $\mathcal{O}(s^3)$ and $\mathcal{O}(s^2)$ coupling-constant sum rules given in Ref.~\cite{Chivukula:2020hvi} {\it do not} hold in the presence of the brane-localized curvature terms. One of the immediate consequences of such violations of these sum rules is that the mass ratio of the second and first (massive) KK modes is no longer bounded by $2$, in contrast to the situation in RS1~\cite{Bonifacio:2019ioc}. See Fig.~\ref{fig:masses}.

\subsection{Feynman gauge and Goldstone boson equivalence theorem \label{subsec:Equivalance}}

The behavior of the scattering amplitudes of massive spin-2 particles is obscure in unitary gauge. Diagrammatically, individual contributions to the amplitudes grow as fast as ${\cal O}(s^5)$ \cite{Arkani-Hamed:2002bjr,Arkani-Hamed:2003roe}, but there are substantial cancellations between the diagrams \cite{SekharChivukula:2019yul,SekharChivukula:2019qih,Chivukula:2020hvi,Chivukula:2021xod,Bonifacio:2019ioc}. In the RS model without brane-localized curvature terms, the analysis is much simpler in `t-Hooft-Feynman gauge. In particular, Ward identities related to the diffeomorphism invariance of the theory relate the amplitudes of the helicity-0 massive spin-2 states to those of the corresponding unphysical spin-0 Goldstone modes \cite{Chivukula:2023qrt,Hang:2021fmp,Hang:2022rjp,Hang:2024uny}. The power-counting of the scattering amplitudes of the spin-0 Goldstone bosons is transparent in this gauge, and all contributions grow no faster than $\mathcal{O}(s)$ -- leading to an overall amplitude which grows (no faster than) this rate. In the presence of the brane-localized curvature terms the power counting in `t-Hooft-Feynman gauge changes, as we now show.

 First we note that, even in the presence of brane-localized curvature interactions, we are still allowed to go to `t-Hooft-Feynman gauge (Sec.~\ref{sec:GF}), where all the Goldstone bosons and the KK gravitons have degenerate spectra. The gauge-fixing terms, when expressed in terms of the mixing of the individual KK modes, are the same whether or not brane-localized curvature interactions are present (leading to the mass-diagonal gauge-fixed kinetic energy terms shown in 
Eq.~(\ref{eq:gauge-fixed-ke})). Therefore the Ward identities derived in \cite{Chivukula:2023qrt} still hold,
\begin{equation}
    \mathcal{M}\left[h^{(n)}_L\Phi\right] = \mathcal{M}\left[\pi^{(n)}\Phi\right] -i\sqrt{3} \mathcal{M}\left[\tilde{A}^{(n)}_L\Phi\right] + \mathcal{M}\left[\tilde{h}^{(n)}_L\Phi\right],
    \label{eq:ward}
\end{equation}
where $\Phi$ represent an ensemble of particles. $\tilde{A}^{(n)}_L$ and $\tilde{h}^{(n)}_L$ are the KK vector Goldstone boson and KK graviton contracted with the modified vector and tensor polarizations,
\begin{equation}
    \tilde{\epsilon}_L^\mu \equiv -\frac{m_n}{E+|\mathbf{p}|} (1, - \mathbf{p}/|\mathbf{p}|)\sim \mathcal{O}\left(\dfrac{m_n}{E}\right),\quad {\rm and}\quad \tilde{\epsilon}_L^{\mu\nu} \equiv \sqrt{\frac{3}{2}}\tilde{\epsilon}_0^\mu\tilde{\epsilon}_0^\nu \sim\mathcal{O}\left(\frac{m_n^2}{E^2}\right).
\end{equation}

However, unlike in the model without brane-localized curvature, where each scalar vertex from the Einstein-Hilbert action can contain at most two derivatives $\partial_\mu$, in the presence of the brane-localized curvature terms each vertex can have more than two derivatives coming from the auxiliary field $\partial_\mu\partial_\nu \Delta$ terms in the metric, Eq.~(\ref{eq:P3}). As we argued, the bulk value of $\Delta$ is unconstrained, and thus the overall scattering amplitudes of physical states must be independent of the bulk value of $\Delta$. Unfortunately, these new auxiliary field contributions reintroduce anomalous high-energy growth diagram by diagram even in `t-Hooft-Feynman gauge. For example, the individual Feynman diagram for $\pi^{(n)}\pi^{(n)}\rightarrow \pi^{(n)}\pi^{(n)}$ contributes terms in the scattering amplitude which grow like $\mathcal{O}(s^5)$ for generic choices of the bulk auxiliary field $\Delta$, and the cancellation of the ${\cal O}(s^5,s^4)$ terms only happens once all the diagrams are summed together. 

Furthermore the last two terms on the right hand side of Eq.~(\ref{eq:ward}) also contribute at $\mathcal{O}(s^3)$ for a generic choice of the auxiliary field $\Delta$, and are not subleading relative to the first term after cancellation, rendering the Ward identity ineffective in isolating the contributions to the scattering amplitude which grow in energy. In the sense that the leading-order contributions to the scattering amplitude are not entirely contained in the scalar Goldstone boson contributions in Eq.~(\ref{eq:ward}) we see that the naive Goldstone boson equivalence theorem (which relates the scattering amplitudes of the helicity-zero spin-2 states to those of the unphysical scalars) fails for generic choices of the auxiliary field $\Delta$.

To avoid the anomalous energy-growth of individual contributions and restore the Goldstone boson equivalence theorem, we must eliminate the bulk part of $\Delta$. We will do so by separating the boundary point from the bulk and set $\Delta = 0 $ in the bulk. We must do so carefully, however, since the derivative $\partial_z$ cannot be defined on the isolated boundary points. Instead, we can define $\Delta$ and its derivatives on the interval $[z_1, z_1+\varepsilon]\cup[z_2-\varepsilon,z_2]$ where $\Delta'$ smoothly goes to zero at $z=z_1+\varepsilon$ and $z_2-\varepsilon$, and set $\Delta = 0$ for $z\in[z_1+\varepsilon, z_2-\varepsilon]$. Then, once we take the limit of $\varepsilon\rightarrow 0^+$, $\Delta$ becomes the distribution,
\begin{equation}
    \Delta = 0,\quad \Delta' = \begin{cases}
        -\dfrac{2\gamma_i}{\theta_i-2A'\gamma_i}\varphi & z =z_i\\
        0 & z_1<z<z_2
    \end{cases},
    \quad
    \Delta'' = \sum_i\delta(z-z_i)\dfrac{2\theta_i\gamma_i}{\theta_i-2A'\gamma_i}\varphi~.
    \label{eq:P3_delta}
\end{equation}

Now that we have eliminated the unphysical contributions arising from the bulk value of $\Delta$, the energy power counting becomes transparent again. The 3-point vertices that have the highest order of four-momenta have the form of
\begin{equation}
    \int dz~Q(z)\left(\partial_z \partial_\mu \partial_\nu  \Delta\right)\left(\partial_z\partial_\rho \partial_\sigma  \Delta\right) \Delta'' \sim \mathcal{O}(p^4),
\end{equation}
where $Q(z)$ represent some function of $z$, and the two derivatives $\partial_z$ come from the Ricci scalar. All the other terms with four or more 4-derivatives vanish due to Eq.~(\ref{eq:P3_delta}). Similarly, the 4- and higher-point vertices contributes only at $\mathcal{O}(p^4)$. 

At the order of $\mathcal{O}(p^4)$, we find that the cubic Lagrangian is given by,
\begin{eqnarray}
    \mathcal{L}_3 &=~& \kappa\int dz~\dfrac{e^{3A}}{24\sqrt{6}}(\eta^{\mu\rho}\eta^{\nu\sigma}-\eta^{\mu\nu}\eta^{\rho\sigma})\left(\partial_\mu \partial_\nu \Delta'\right)\left(\partial_\rho \partial_\sigma \Delta'\right)\Delta'' + \mathcal{O}(p^3)\\
    &=~& \sum_i\dfrac{\kappa}{6\sqrt{6}}\left[e^{3A}\left(\dfrac{\theta_i\gamma_i}{\theta_i-2A'\gamma_i}\right)^3 \left(\partial_\mu\varphi\right)\left(\partial^\mu\varphi\right)\left(\Box\varphi\right)\right]_{z=z_i} + \mathcal{O}(p^3) \label{eq:L3_1}\\
    &=~& \sum_{n_1,n_2,n_3} \sum_{i=1}^2\dfrac{\kappa}{6\sqrt{6}}\left[e^{3A}\left(\dfrac{\theta_i\gamma_i}{\theta_i-2A'\gamma_i}\right)^3 k^{(n_1)}k^{(n_2)}k^{(n_3)}\right]_{z=z_i} \nonumber\\
    &&~\times \left(\partial_\mu\pi^{(n_1)}\right)\left(\partial^\mu\pi^{(n_2)}\right)\left(\Box\pi^{(n_3)}\right) + \mathcal{O}(p^3),
\label{eq:vertex3}
\end{eqnarray}
where we have used integration by parts to write the interaction in the form that commonly appears in the literature for the Galileon~\cite{Dvali:2000hr,Luty:2003vm,Nicolis:2008in,deRham:2010ik,Hinterbichler:2011tt,deRham:2014zqa}. Note that when we integrate over $z$, one has to be careful: the identity $\int dx f(x)\delta(x)=f(0)$ is true only if $f(x)$ is continuous at $x=0$.

Using the interaction given in Eq.~(\ref{eq:vertex3}) and the Goldstone boson equivalence theorem, one can compute the scattering amplitude of four longitudinally polarized KK gravitons at the order of $\mathcal{O}(s^3)$, by computing the amplitude of four scalar Goldstone bosons,
\begin{equation}
\aligned
    \widetilde{M}^{(6)}[h_L^{(n)}h_L^{(n)}\rightarrow h_L^{(n)} h_L^{(n)}] =~& \widetilde{M}^{(6)}[\pi^{(n)}\pi^{(n)}\rightarrow \pi^{(n)} \pi^{(n)}] \\
    & \hspace{-2cm}=~ -\sum_{j=0}^{\infty}\dfrac{\kappa^2\sin^2\theta}{288}\left\{\sum_{i=1}^{2}\left[e^{3A}\left(\dfrac{\theta_i\gamma_i}{\theta_i-2A'\gamma_i}\right)^3 k^{(n)}k^{(n)}k^{(j)}\right]_{z=z_i}\right\}^2~.
\endaligned
\end{equation}
We have checked numerically that these partial amplitudes reproduce the results found in unitary gauge in the computations described in the previous subsection.
Note that these contributions vanish in the limit $\gamma_{1,2}\to 0$, as they must since we know that the amplitudes in that case grow only as fast as ${\cal O}(s)$.

Thus we have shown that, in the presence of the brane-localized curvature terms, the leading contributions to the scattering amplitudes of helicity-0 massive spin-2 KK modes can be understood as arising from the Galileon interactions of the tower of corresponding spin-0 Goldstone bosons. 
It is these interactions that are responsible for the new ${\cal O}(s^3)$ growth of the elastic scattering amplitudes of helicity-0 massive spin-2 KK modes. This tower of Galileon interactions for the spin-0 Goldstone modes are the result of the auxiliary field $\Delta$, which was required because of the incompatibility between the straight gauge condition $\xi^5=0$ and the non-Dirichlet condition of $g^{(n)}$ given in Eq.~(\ref{eq:bc_fgk}). 

Before ending this subsection, we would like to briefly comment on the computation of the subleading contributions. The downside of choosing $\Delta$ to be a distribution in Eq.~(\ref{eq:P3_delta}) is that, at subleading orders $\mathcal{O}(s^{2})$, one will encounter overlap integrals naively of the form,
\begin{equation}
    \int dz~Q(z)\delta(z-z_i)\delta(z-z_i).
\end{equation}
One would need to carefully regulate such divergences by defining the distributions correctly, such as treating them as the limit of some continuous functions
and showing that the final results were finite in the appropriate limit.

\subsection{The Strong Coupling Scale $\Lambda_3$}

In the presence of brane-localized curvature interactions, the ${\cal O}(s^3)$ high-energy behavior of the scattering amplitudes implies the presence of a new scale, $\Lambda_3$, which characterizes when the corresponding partial amplitude 
saturates unitarity. In this section we describe how this new parameter compares to other dimensionful scales in the theory.

In general, we define the scale associated with a partial amplitude which grows with energy as the scale at which its contribution to the (properly normalized) scattering amplitude would saturate the unitarity bound. For the 2-to-2 scattering with helicities $(\lambda_1,\lambda_2)\rightarrow(\lambda_3,\lambda_4)$, one can derive the partial wave amplitudes
\begin{equation}
    a^J=\dfrac{1}{32\pi^2}\int d\cos\theta d\phi ~D^J_{\lambda_1-\lambda_2,\lambda_3-\lambda_4}(\theta,\phi)\mathcal{M}(s,\theta,\phi),
\end{equation}
where $D^J_{\lambda,\lambda'}(\theta,\phi)$ is the Wigner D function with the normalization
\begin{equation}
    \int d\cos\theta d\phi ~\left|D^J_{\lambda,\lambda'}(\theta,\phi)\right|^2 = \dfrac{4\pi}{2J+1}.
\end{equation}
The unitarity bounds at large $s$ are derived by
\begin{equation}
    \left|{\rm Re}(a^J)\right| \le \dfrac{1}{2}.
\end{equation}

Absent brane-localized curvature terms the RS model is specified by three parameters: the Planck mass $M_{\rm Pl}$, the hierarchy $z_2/z_1=e^{k\pi r_c}$, and the mass of the lowest massive spin-2 KK meson $m_1$. For fixed $M_{\rm Pl}$ and $m_1$, we define the ``RS" scale in terms of the unitarity bound on the scattering process $h_L^{(n)}h_L^{(n)}\rightarrow h_L^{(n)} h_L^{(n)}$, and find 
\cite{Chivukula:2020hvi, Chivukula:2023qrt}
\begin{equation}
    \Lambda_{\rm RS} = \left(\dfrac{\kappa^2}{8\pi} \int_{z_1}^{z_2} dz~e^{3A}\left[k^{(n)}(z)\right]^4\right)^{-1/2},
\end{equation}
where the relation between $\kappa$ and the 4D Planck scale $M_{\rm Pl}$ is given by
\begin{equation}
    \kappa^2 M_{\rm Pl}^2 = \dfrac{2}{k}\left[1-\dfrac{1}{(k z_2)^2}\right].
\end{equation}
This scale is typically of order the Planck scale ``warped down" by a factor of $z_1/z_2 = e^{-k\pi r_c}$.

In the presence of brane-localized curvature terms, the unitarity bound on the partial amplitude $\widetilde{M}^{(6)}$ defines a new scale
\begin{equation}
    \Lambda_3 = \left(\dfrac{\kappa^2}{1728\pi} \sum_{j=0}^{\infty}\left\{\sum_{i=1}^{2}\left[e^{3A}\left(\dfrac{\theta_i\gamma_i}{\theta_i-2A'\gamma_i}\right)^3 k^{(n)}k^{(n)}k^{(j)}\right]_{z=z_i}\right\}^2\right)^{-1/6},
\end{equation}
where the relation between $\kappa$ and the 4D Planck scale $M_{\rm Pl}$ is given by
\begin{equation}
    \kappa^2 M_{\rm Pl}^2 = \dfrac{2}{k}\left[1-\dfrac{1}{(k z_2)^2}+2\gamma_1 k + \dfrac{2 \gamma_2}{k^2 z_2^3}\right].
\end{equation}

\begin{figure}[tbh]
    \centering
    \includegraphics[width=0.48\textwidth]{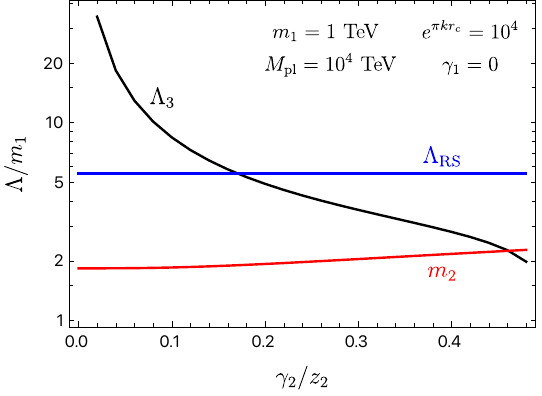}
    \includegraphics[width=0.48\textwidth]{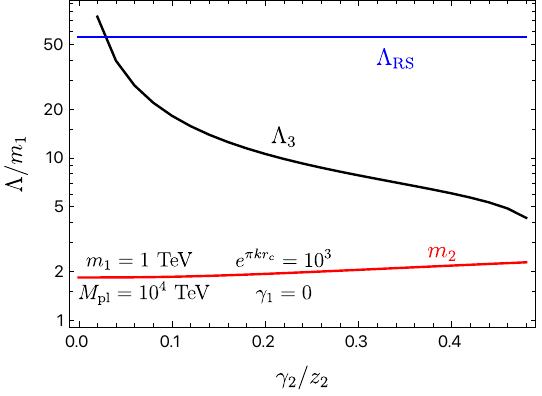}
    \caption{The unitarity scales derived for the scattering of the  KK modes $h_L^{(1)}h_L^{(1)}\rightarrow h_L^{(1)} h_L^{(1)}$. All energy scales are given in terms of $m_1$; for the purposes of illustration, we take $m_1 = 1$ TeV and with $M_{\rm Pl} = 10^4$ TeV. The three energy scales plotted are $\Lambda_3$, $\Lambda_{\rm RS}$, and the mass of the second spin-2 KK state $m_2$. Since $\gamma_1$ has little numerical impact on the scales shown, it is taken to be zero in both graphs. The variation of the scales is shown as a function of $\gamma_2/z_2$ over the allowed region. In both graphs we see that $\Lambda_3 < \Lambda_{\rm RS}$ and $m_2/m_1$ exceeds two for larger values of $\gamma_2/z_2$. Left: $z_2/z_1 = 10^4$. Right: $z_2/z_1=10^3$.  }
    \label{fig:scales}
\end{figure}

A comparison of the scales $\Lambda_3$ and $\Lambda_{\rm RS}$ is shown in Fig.~\ref{fig:scales}. In both cases we see that the effect of the brane-localized curvature interactions is, as expected, enhanced as $\gamma_2/z_2$ grows. In particular, as the brane-localized curvature interaction becomes larger, $\Lambda_3$ drops below $\Lambda_{\rm RS}$ -- and therefore sets the range of validity of the effective KK theory. We also see again that the ratio $m_2/m_1$ of the second spin-2 KK resonance mass to the first grows and is larger than 2.

\section{DGP model and the $\Lambda_3$ scale}
\label{sec:DGP}

The Dvali-Gabadadze-Porrati (DGP) model~\cite{Dvali:2000hr} is an extra-dimensional model which has received considerable attention as an alternative description of our accelerating universe. It is also one of the first ghost-free examples of the localization of gravity on a 4D brane in a semi-infinite transverse space. In this section, we show how the continuum of massive gravitons and the strong interaction scale in the DGP model~\cite{Luty:2003vm} can be derived using our formalism.\footnote{For a review in the context of massive gravity, see \cite{Hinterbichler:2011tt,deRham:2014zqa}.}

The DGP model has only one brane embedded in a flat fifth dimension, and contains 5D gravity in the bulk and 4D gravity localized on the brane. In our formalism, the DGP model can be realized by taking the second brane to infinity and setting the warp factor to zero,
\begin{equation}
    A(z) = 0,\quad z_1=0, \quad z_2\rightarrow \infty,\quad \gamma\equiv\gamma_1,\quad \gamma_2 = 0,
\end{equation}
where we also set $z_1=0$ for convenience.
Since the fifth dimension is semi-infinite, there is no KK decomposition into discrete modes. Instead, we impose the boundary conditions $H_{MN}(z_2)\rightarrow 0$ as $z_2\rightarrow\infty$, and the spectrum of the massive gravitons become a continuum. 

Using the metric decomposition in Eq.~(\ref{eq:background-metric}) and setting $A(z)=0$,  the gauge-fixed quadratic terms in the 't~Hooft-Feynman gauge\footnote{The gauge-fixing terms chosen in \cite{Luty:2003vm} correspond to this gauge.} become (see Eqs.~(\ref{eq:RS1Lagrangian}-\ref{eq:RS1phiphi}) and Eqs.~(\ref{eq:tHooft-gauge}-\ref{eq:F5}))
\begin{equation}
\aligned
    S =& \int d^4x\int_{z_1}^{\infty}dz~\left\{\vphantom{\dfrac{1}{4}}\right.\dfrac{1}{4}\left(\eta^{\mu\rho}\eta^{\nu\sigma} + \eta^{\mu\sigma}\eta^{\nu\rho} - \eta^{\mu\nu}\eta^{\rho\sigma}\right)\left[-h_{\mu\nu}\Box h_{\rho\sigma} \left(1+\gamma\delta(z)\right) - \left(\partial_5 h_{\mu\nu}\right)\left(\partial_5 h_{\rho\sigma}\right)\right] \\
    &\hspace{2cm} + \dfrac{1}{2}\left(-\eta^{\mu\nu}\right)\left[-A_{\mu}\Box A_{\nu} \left(1+\gamma\delta(z)\right) - \left(\partial_5  A_{\mu}\right)\left(\partial_5 A_{\nu}\right)\right] \\
    &\hspace{2cm} + \dfrac{1}{2}\left[-\varphi\Box\varphi \left(1+\gamma\delta(z)\right) - \left(\partial_5\varphi\right)^2\right]\left.\vphantom{\dfrac{1}{2}}\right\}~.
\endaligned
\end{equation}
Since the bulk equations ($z>0$) of all of the fields are of form
\begin{equation}
    (\Box-\partial_z^2)\Phi = 0,
\end{equation}
where $\Phi = h_{\mu\nu}$, $A_\mu$, and $\varphi$, we follow the usual analysis \cite{Dvali:2000hr,Luty:2003vm,Nicolis:2008in,deRham:2010ik,Hinterbichler:2011tt,deRham:2014zqa} and write the generic solutions (subject to the boundary condition $H_{MN}\to 0$ as $z\to \infty$) in terms of nonlocal wavefunctions,
\begin{equation}
    \Phi(x,z) = \dfrac{1}{\sqrt{\gamma}}e^{-z\sqrt{\Box}}\tilde{\Phi}(x),
\end{equation}
where the normalization factor is taken for convenience. Defining the bulk fields in terms of their boundary values is the analog of the Kaluza-Klein expansion in Eqs.~(\ref{eq:KK_1u} - \ref{eq:KK_3u}). Plugging this form into the action and (formally) performing the integration over $z$ we obtain action in terms of the boundary-value fields
\begin{equation}
\aligned
    S =&~\int d^4x~\left\{\vphantom{\dfrac{1}{4}}\right.\dfrac{1}{4}\left(\eta^{\mu\rho}\eta^{\nu\sigma} + \eta^{\mu\sigma}\eta^{\nu\rho} - \eta^{\mu\nu}\eta^{\rho\sigma}\right)\left[\tilde{h}_{\mu\nu}\left(-\Box - \dfrac{1}{\gamma}\sqrt{\Box} \right) \tilde{h}_{\rho\sigma} \right] \\
    &\hspace{1cm} + \dfrac{1}{2}\left(-\eta^{\mu\nu}\right)\left[\tilde{A}_{\mu}\left(-\Box - \dfrac{1}{\gamma}\sqrt{\Box} \right) \tilde{A}_{\nu}\right] \\
    &\hspace{1cm} + \dfrac{1}{2}\left[\tilde{\varphi}\left(-\Box - \dfrac{1}{\gamma}\sqrt{\Box} \right)\tilde{\varphi} \right]\left.\vphantom{\dfrac{1}{2}}\right\}.
\endaligned
\end{equation}
Having performed the integral over the extra dimension the kinetic energy terms come from the brane-localized curvature interactions and, in the 't Hooft-Feynman gauge, all the fields have the same continuous spectrum. Here the non-local $\sqrt{\Box}$ operator is to be understood via its Fourier transform 
\cite{Dvali:2000hr,Luty:2003vm,Nicolis:2008in,deRham:2010ik,Hinterbichler:2011tt,deRham:2014zqa}, and the corresponding propagator has a ``soft mass" proportional to $1/\gamma$.
%\begin{equation}
%\aligned
%    \int_{z_1}^\infty dz~\mathcal{L}_{\varphi\mbox{-}\varphi,\,\rm gauge\mbox{-}fixed} =&~ -\int_{z_1}^\infty dz~\left[\frac{1}{2}\varphi\Box\varphi \left(1+\gamma\delta(z-z_1)\right) + \dfrac{1}{2}\left(\overline{D}^\dagger\varphi\right)^2\right]\\
%    =&~ -\dfrac{1}{2}\tilde{\varphi}\left(\Box + \dfrac{1}{\gamma}\sqrt{\Box} \right)\tilde{\varphi}
%\endaligned
%\end{equation}.

We may now directly apply our analysis of the leading interactions of the scalar fields $\varphi$, which we have seen also come from the brane-curvature terms. In particular, the cubic interaction in Eq.~(\ref{eq:L3_1}) becomes,
\begin{equation}
    \mathcal{L}_3 = \dfrac{\kappa\gamma^{3/2}}{6\sqrt{6}}(\partial_\mu\tilde{\varphi})(\partial^\mu\tilde{\varphi})(\Box\tilde{\varphi}).
\end{equation}
Translating into the notation in Ref.~\cite{Luty:2003vm},
\begin{equation}
    \kappa = \frac{\sqrt{2}}{M_5^{3/2}},\qquad \gamma=\dfrac{M_4^2}{M_5^3},
\end{equation}
we recover the effective Galileon interaction among the scalar Goldstone bosons and the DGP strong interaction scale~\cite{Luty:2003vm}
\begin{equation}
    \mathcal{L}_3 = \dfrac{1}{6\sqrt{3}\Lambda_3^3}(\partial_\mu\tilde{\varphi})(\partial^\mu\tilde{\varphi})(\Box\tilde{\varphi}),\qquad \text{with}~\Lambda_3 = \left(\dfrac{\kappa\gamma^{3/2}}{\sqrt{2}}\right)^{-1/3} = \dfrac{M_5^2}{M_4}.
\end{equation}

\section{Conclusion}
\label{sec:conclusion}

In this paper we have presented an investigation of the properties of the scattering amplitudes of the massive spin-2 Kaluza-Klein states of extra-dimensional theories of gravity in the presence of brane-localized curvature interactions. We have shown that the presence of these new interactions modifies the high-energy behavior of these amplitudes from ${\cal O}(s)$ to ${\cal O}(s^3)$, spoiling some of the ``sum-rule" relationships between the masses and couplings of the massive spin-2 states which are present in theories without these interactions. We have explained how the scale $\Lambda_3$ related to the ${\cal O}(s^3)$ growth is related to the intrinsic gravitational scale $\Lambda$ and the spectrum of spin-2 Kaluza-Klein states, in terms of the parameters defining the theory. Using the diffeomorphism invariance of the theory, and the hidden supersymmetric structure of the Sturm-Liouville systems associated with the Kaluza-Klein mode systems, we have demonstrated that the there are Galileon interactions of the scalar modes of the massive spin-2 tower which explain the behavior of the high-energy scattering amplitudes. We have studied the gauge-invariance of the theory, and shown that gauge-invariance depends crucially on identifying how diffeomorphism transformations act on the metric fluctuations of the compactified theory. We have described what happens to our results for a finite extra dimension in the limit in which we move the UV brane to infinity, giving a natural and gauge-invariant explanation of the properties of scattering in the DGP model.

Next, we briefly comment on the extension of our results to a model \cite{George:2011sw} in which the size of the extra dimension is stabilized, and in which the radion is no longer massless, via the incorporation of a Goldberger-Wise (GW) mechanism \cite{Goldberger:1999uk,Goldberger:1999un}. While a detailed analysis of the properties of a GW-stabilized RS model in the presence of brane-localized curvature interactions is beyond the scope of this work, we expect that the extension of the results given here to that case should be straightforward. In particular, it has previously been shown that the modified boundary conditions of the GW model (which mix metric and bulk scalar fluctuations) {\it without} brane-localized curvature interactions do still allow for the definition of a SUSY structure for the mode eigensystems and a corresponding understanding of diffeomorphism invariance \cite{Chivukula:2022kju} -- resulting in scattering amplitudes which grow only as fast as ${\cal O}(s)$. The new ingredient in the presence of brane-localized curvature interactions is the necessity of the auxiliary field. However, as pointed out in Ref.~\cite{George:2011sw}, the boundary constraints on the auxiliary field $\Delta$ in Eq.~(\ref{eq:P3}) used here {\it are unchanged} by the GW stabilizing interactions themselves. Therefore we expect that it is possible to generalize the results of \cite{Chivukula:2022kju} to include brane-localized curvature terms, and we believe that the high-energy behavior of the massive spin-2 scattering amplitudes will continue to be of order ${\cal O}(s^3)$.

 The phenomenological implications of the results obtained here are potentially wide-ranging. Since we are ignorant of the UV gravity dynamics, it is quite plausible that brane-localized curvature terms are produced generically in many such theories. The resulting ${\cal O}(s^3)$ growth, the corresponding $\Lambda_3$ cutoff, has consequences in a variety of phenomenological scenarios, including electroweak phase-transitions in the early universe\cite{Dillon:2017ctw}, flavor physics \cite{Agashe:2016rle}, electroweak symmetry breaking \cite{Davoudiasl:2005uu}, gravitational wave probes \cite{Bouhmadi-Lopez:2004ruo} as well as gravity mediated supersymmetry breaking \cite{Rattazzi:2003rj} to name a few. In subsequent work, we will analyze the phenomenological consequences of our finding.

%%%%%%%%%%%%%%%%%%%%%%%%%%%%%%%%%%%
\section*{Acknowledgements}
We thank Rong-Xin Miao for bringing ref. \cite{Miao:2023mui} to our attention.

RSC, EHS, and XW were supported, in part, by by the US National Science Foundation under Grant No. PHY-2210177.
The work of KM was supported in part by the National Science Foundation under Grant No. PHY-2310497. DS is partially supported through the the University of New South Wales, Sydney startup grant PS-71474 and in part by the the Australian Research Council through the Centre of Excellence for Dark Matter Particle Physics (CE200100008).

%%%%%%%%%%%%%%%%%%%%%%%%%%%%%

\appendix

\section{Gauge theory with brane-localized kinetic terms in the Randall-Sundrum background }

\label{sec:appendix}

In this appendix we describe how to gauge-fix a compactified 5-dimensional gauge theory in the presence of brane gauge kinetic energy terms and preserve the $N=2$ SUSY structure which ensures (a subset of) five-dimensional gauge-invariance. We show how the presence of brane-localized gauge kinetic-energy interactions does modify the form of the gauge-transformations of the compactified gauge theory.  This material generalizes the construction given in  \cite{Lim:2005rc, Lim:2007fy}.

Consider a gauge theory in the bulk with brane-localized kinetic terms,
\begin{equation}
    S = \int d^4x \left[\int_{z_1}^{z_2}dz~\sqrt{G}\left(-\dfrac{1}{4}F^{MN}F_{MN}\right) + \left.\sum_{i=1,2}\gamma_i \sqrt{\bar{G}}\left(-\dfrac{1}{4}F^{\mu\nu}F_{\mu\nu}\right)\right|_{z=z_i}\right].
\end{equation}
We discuss an Abelian gauge-theory here explicitly for convenience, but the mode equations and the $N=2$ SUSY properties of an non-Abelian theory are identical.
The quadratic terms in the Lagrangian can be written as 
\begin{equation}
\aligned
    \mathcal{L}_{VV} = &~ \dfrac{1}{2}\left\{\left[\int_{z_1}^{z_2}dz~e^{A(z)}V^\mu\left(\eta_{\mu\nu}\partial_\rho\partial^\rho - \partial_\mu\partial_\nu \right)V^\nu\right] + \left.\sum_i\gamma_i~e^{A(z)}V^\mu\left(\eta_{\mu\nu}\partial_\rho\partial^\rho - \partial_\mu\partial_\nu \right)V^\nu\right|_{z=z_i}\right.\\
    &~ \left. + \int_{z_1}^{z_2}dz~e^{A(z)} \left(\partial_z V_\mu\partial_z V^\mu - V_5\partial_\mu\partial^\mu V_5 + 2V_5\partial_\mu \partial_zV^\mu\right)\right\}.
\endaligned
\end{equation}
We expand the KK modes as \cite{Lim:2005rc}
\begin{eqnarray}
    V_\mu(x^\alpha,z) &=& \sum_n V_\mu^{(n)}(x^\alpha)f^{(n)}_{V}(z),\\
    V_5(x^\alpha,z) &=& \sum_n V_5^{(n)}(x^\alpha)f^{(n)}_{V_5}(z).
\end{eqnarray}
To make the KK Lagrangian kinetic terms canonical, we require that the wavefunctions satisfy
\begin{eqnarray}
    &&\left(\int_{z_1}^{z_2}dz~e^{A(z)} f_V^{(m)}(z)f_V^{(n)}(z)\right) + \left.\sum_i \gamma_i~e^{A(z)} f_V^{(m)}(z)f_V^{(n)}(z)\right|_{z=z_i} = \delta_{mn},
    \label{eq:gauge-inner-product}\\
    && \int_{z_1}^{z_2}dz~e^{A(z)} \partial_z f_V^{(m)}(z) \partial_zf_V^{(n)}(z) = m_{V,n}^2\delta_{mn},\\
    && \int_{z_1}^{z_2}dz~e^{A(z)} f_{V_5}^{(m)}(z)f_{V_5}^{(n)}(z) = \delta_{mn}.
\end{eqnarray}

The SUSY relations are as usual~\cite{Lim:2005rc,Lim:2007fy,Chivukula:2023sua},
\begin{equation}
    \begin{cases}
        D_V f^{(n)}_{V} = m_{V, n}f^{(n)}_{V_5}~ \\
        D_V^\dagger f^{(n)}_{V_5} = m_{V, n}f^{(n)}_{V}~
    \end{cases}.
\end{equation}
where
\begin{equation}
    D_V =\partial_z, \qquad
    D_V^\dagger = -\partial_z -A'(z).
\end{equation}
The only twist is the additional  change in the definition of the inner product in a manner analogous to Eqs.~(\ref{eq:h-inner-product})-(\ref{eq:k-inner-product}),
\begin{eqnarray}
    \braket{f^{(m)}_V | f^{(n)}_V} &=& \left(\int_{z_1}^{z_2}dz~e^{A(z)} f_V^{(m)}(z)f_V^{(n)}(z)\right) + \left.\sum_i \gamma_i~e^{A(z)} f_V^{(m)}(z)f_V^{(n)}(z)\right|_{z=z_i},\\
    \braket{f^{(m)}_{V_5} | f^{(n)}_{V_5}} &=& \left(\int_{z_1}^{z_2}dz~e^{A(z)} f_{V_5}^{(m)}(z)f_{V_5}^{(n)}(z)\right).
\end{eqnarray}

To construct the SUSY algebra, we define
\begin{equation}
    \Psi = \begin{pmatrix} f_V \\ f_{V_5} \end{pmatrix},
\end{equation}
and the inner product
\begin{equation}
\aligned
    \braket{\tilde{\Psi} | \Psi} =&~ \left(\int_{z_1}^{z_2}dz~e^{A} \tilde{f}_Vf_V\right) + \left.\sum_i \gamma_i~e^{A} \tilde{f}_V(z)f_V(z)\right|_{z=z_i} + \left(\int_{z_1}^{z_2}dz~e^{A} \tilde{f}_{V_5}f_{V_5}\right)
\endaligned
\end{equation}
Now we can define the supercharge $Q$ and $Q^\dagger$,
\begin{equation}
    Q = \begin{pmatrix} 0&0\\D_V&0 \end{pmatrix},\qquad Q^\dagger = \begin{pmatrix} 0&D_V^\dagger\\0&0 \end{pmatrix}.
\end{equation}

In order for the boundary conditions to respect the supersymmetry as well, the supercharges $Q$ and $Q^\dagger$ must be the hermitian conjugates of each other with respect to the inner product,
\begin{equation}
    \braket{\tilde{\Psi}|Q\Psi} = \begin{pmatrix} \tilde{f}_V & \tilde{f}_{V_5}\end{pmatrix} \begin{pmatrix} 0&0\\D_V&0 \end{pmatrix} \begin{pmatrix} f_V \\ f_{V_5} \end{pmatrix} =  \int_{z_1}^{z_2} dz e^A \tilde{f}_{V_5}D_Vf_V,
\end{equation}
\begin{equation}
\aligned
    \braket{Q^\dagger\tilde{\Psi}|\Psi} =&~ \left(\int_{z_1}^{z_2}dz~e^{A} f_VD_V^\dagger \tilde{f}_{V_5}\right) + \left.\sum_i \gamma_i~e^{A} f_VD_V^\dagger \tilde{f}_{V_5}\right|_{z=z_i} \\
    =&~ \left(\int_{z_1}^{z_2}dz~e^{A} \tilde{f}_{V_5} D_Vf_V\right) + \left.e^{A} f_V\tilde{f}_{V_5}\right|_{z=z_2} - \left.e^{A} f_V\tilde{f}_{V_5}\right|_{z=z_1} \\
    &~ + \left.\gamma_1~e^{A} f_VD_V^\dagger \tilde{f}_{V_5}\right|_{z=z_1} + \left.\gamma_2~e^{A} f_VD_V^\dagger \tilde{f}_{V_5}\right|_{z=z_2}
\endaligned
\end{equation}
Thus, by requiring
\begin{equation}
    \braket{\tilde{\Psi}|Q\Psi} = \braket{Q^\dagger\tilde{\Psi}|\Psi},
\end{equation}
one can derive the boundary conditions\footnote{There exist other solutions that contain either $f_V(z_1)=0$ or $f_V(z_2)=0$. They are less interesting because they would lead to vanishing brane kinetic terms at either $z=z_1$ or $z_2$. Also, at least in the case of $\gamma_1=\gamma_2=0$, these solutions lead to scenarios where either $V_\mu$ has no massless mode or $V_5$ has a physical massless mode.}
\begin{eqnarray}
    \left.f_{V_5}  - \gamma_1 D_V^\dagger f_{V_5}\right|_{z=z_1} = \left.f_{V_5}  + \gamma_2 D_V^\dagger f_{V_5}\right|_{z=z_2} = 0.
\end{eqnarray}
Note that the hermiticity requirement does not put constrains on $f_V$, thus the above boundary conditions do not conflict with the SUSY relation. And instead, the SUSY relation,
\begin{equation}
    D_V f^{(n)}_{V} = m_{V, n}f^{(n)}_{V_5},
\end{equation}
leads us to the boundary conditions for $f_V$,
\begin{equation}
    \left.D_Vf_V  - \gamma_1 D^\dagger D_Vf_{V}\right|_{z=z_1} = \left.D_Vf_{V}  + \gamma_2 D^\dagger D_Vf_{V}\right|_{z=z_2} = 0.
\end{equation}
When written in terms of a specific mode function, analogous to the form shown in Eqs.~(\ref{eq:spin2-eq1}) and (\ref{eq:spin2-eq2}), the second derivatives can be recast in terms of the mass eigenvalues $m^2_n$ and interpreted as a Sturm-Liouville system with respect to the weight function of Eq.~(\ref{eq:gauge-inner-product}).

In terms of these modes we find the subset of gauge transformations respected by the compactified theory have parameters that may be written\footnote{We write the transformation here for a $U(1)$ gauge theory}
\begin{align}
    \Theta(x^\mu,z) & = \sum_n \theta^{(n)}(x^\mu) f^{(n)}_{V}(z)~.
\end{align}
Under these gauge transformations we find \cite{Lim:2005rc}
\begin{align}
    V_M(x^\alpha,z) & \to V_M(x^\alpha, z) + \partial_M \Theta(x^\alpha,z)~,\label{eq:gauge-trans1}\\
    V_\mu^{(n)}(x^\mu) & \to  V_\mu^{(n)}(x^\mu) + \partial_\mu \theta^{(n)}(x^\alpha)~, \\
    V_5^{(n)}(x^\alpha) & \to V_5^{(n)}(x^\alpha) + m_n \theta^{(n)}(x^\alpha)~. \label{eq:gauge-trans2}
\end{align}
The $R_\xi$ gauge-fixing terms are then given by
\begin{equation}
    S_{\rm GF} = -\dfrac{1}{2\xi}\int d^4x\left\{
    \left[\int_{z_1}^{z_2}dz~e^{A}\left(\partial_\mu V^\mu - \xi D_V^\dagger V_5\right)^2\right] + 
    \left.\sum_i\gamma_i~e^{A}\left(\partial_\mu V^\mu - \xi D_V^\dagger V_5\right)^2\right|_{z=z_i}\right\}.
\end{equation}

While the expressions given here are for an Abelian theory, the generalization to a non-Abelian theory are straightforward by introducing gauge-field components, gauge parameters, and gauge-fixing functions for each generator, and introducing appropriate {\it homogeneous} transformation terms in Eqs.~(\ref{eq:gauge-trans1})-(\ref{eq:gauge-trans2}).
The usual Goldstone boson equivalence theorem relationships in `t-Hooft-Feynman gauge ($\xi=1$) between the scattering amplitudes of the longitudinal massive spin-1 KK states and the corresponding Goldstone bosons ($V_5^{(n)}(x^\alpha)$) then follow from the discussion above, and result in the high-energy behavior for the scattering amplitudes found previously \cite{Chivukula:2001esy}. 

Note that in the case of gauge theory there are no background fields, and therefore no auxiliary field is required to define the theory. Hence the presence of brane-localized gauge-kinetic terms does not change the energy dependence of the Goldstone-boson interactions, and massive spin-1 KK scattering amplitudes continue to grow like a constant at high-energies \cite{Chivukula:2001esy}. This is  unlike the case of compactified gravity where, due to the presence of a background metric, brane-localized curvature creates a conflict between the modifications of diffeomorphism invariance and the straight gauge conditions of Eq.~(\ref{eq:straight-gauge}) requiring the inclusion of the auxiliary field and modifying the behavior of the scalar metric fluctuations.

\bigskip

\bibliographystyle{apsrev4-1.bst}

\bibliography{ref}{}

\end{document}